\documentclass[12pt,preprint]{aastex}

\newcommand{\ldot}{L$_{\odot}$}
\newcommand{ \um}{$\mu$m~}
\newcommand{ \ums}{$\mu$m}
\def\kmsMpc{\ifmmode {\rm\,km\,s^{-1}\,Mpc^{-1}}\else
    ${\rm\,km\,s^{-1}\,Mpc^{-1}}$\fi}

\shorttitle{ Classification and Luminosities for Dusty Quasars}
\shortauthors{Weedman et al.}

\begin{document}

\title{Infrared Classification and Luminosities For Dusty AGN and the Most Luminous Quasars}

\author{Daniel Weedman\altaffilmark{1}, Lusine Sargsyan\altaffilmark{1}, Vianney Lebouteiller\altaffilmark{2}, James Houck\altaffilmark{1}, and Donald Barry\altaffilmark{1}}

\altaffiltext {1}{Astronomy Department, Cornell University, Ithaca,
NY 14853; dweedman@isc.astro.cornell.edu}
\altaffiltext {2} {Laboratoire AIM, CEA/DSM-CNRS-Universite Paris Diderot, DAPNIA/Service d'Astrophysique, Saclay, France}

\begin{abstract}
 
Mid-infrared spectroscopic measurements from the Infrared Spectrometer on $Spitzer$ (IRS) are given for 125 hard X-ray AGN (14-195 keV) from the $Swift$ Burst Alert Telescope sample and for 32 AGN with black hole masses from reverberation mapping. The 9.7 \um silicate feature in emission or absorption defines an infrared AGN classification describing whether AGN are observed through dust clouds, indicating that 55\% of the BAT AGN are observed through dust. The mid-infrared dust continuum luminosity is shown to be an excellent indicator of intrinsic AGN luminosity, scaling closely with the hard X-ray luminosity, log $\nu L_{\nu}$(7.8 \ums)/$L$(X) = -0.31 $\pm$ 0.35 and independent of classification determined from silicate emission or absorption.  Dust luminosity scales closely with black hole mass, log $\nu L_{\nu}$(7.8 \ums) = (37.2 $\pm$ 0.5) + 0.87 log BHM for luminosity in erg s$^{-1}$ and BHM in M$_{\odot}$.  The 100 most luminous type 1 quasars as measured in $\nu L_{\nu}$(7.8 \ums) are found by comparing Sloan Digital Sky Survey optically discovered quasars with photometry at 22 \um from the Wide-Field Infrared Survey Explorer, scaled to rest frame 7.8 \um using an empirical template determined from IRS spectra.  The most luminous SDSS/WISE quasars have the same maximum infrared luminosities for all 1.5 $<$ z $<$ 5, reaching total infrared luminosity $L_{IR}$ = 10$^{14.4}$ \ldot.  Comparing with Dust Obscured Galaxies from $Spitzer$ and WISE surveys, we find no evidence of hyperluminous obscured quasars whose maximum infrared luminosities exceed the maximum infrared luminosities of optically discovered quasars.  Bolometric luminosities $L_{bol}$ estimated from rest frame optical or ultraviolet luminosities are compared to $L_{IR}$.  For the local AGN, the median log $L_{IR}$/$L_{bol}$ = -0.35, consistent with a covering factor of 45\% for the absorbing dust clouds. For the SDSS/WISE quasars, the median log $L_{IR}$/$L_{bol}$ = 0.1, with extremes indicating that ultraviolet-derived $L_{bol}$ can be seriously underestimated even for type 1 quasars.

\end{abstract}


\section{Introduction}

Observational understanding of the initial assembly of galaxies within the early Universe arises by tracing the luminosities and characteristics of galaxies as a function of redshift.  The astrophysical questions are fundamental: How and when did the first generation of stars form and assemble into galaxies?  When and how did the supermassive black holes (SMBH) within quasars or active galactic nuclei (AGNs) develop? How massive can be early galaxies and SMBH? What is the connection between initial star formation and AGN?

A specific observational challenge is to trace these formation processes to their highest redshifts.  For luminous quasars and luminous star formation, this has been most successful using optically discovered sources whose luminosities seem to peak at 2 $\la$ z $\la$ 3 when observed in the rest frame ultraviolet \citep[e.g.][]{mad98,fan04,cro04,red09}.  It is also known, however, that there are many luminous but dusty sources whose dust obscures the primary optical, ultraviolet and even X-ray sources of luminosity.  The existence of such dusty objects means that a census of the universe derived only from optical observations must be incomplete.

It is known, for example, that the most luminous galaxies ($L$ $>$ 10$^{13}$ \ldot) in the local universe are the Ultraluminous Infrared Galaxies (ULIRGs, e.g. Soifer, Neugebauer and Houck 1987, Sanders and Mirabel 1996) whose luminosity arises from reemission by dust at infrared wavelengths of electromagnetic radiation initially generated at much shorter wavelengths.  Surveys in the submillimeter initially discovered individual, optically obscured, dusty sources at redshifts z $\ga$ 2 \citep{cha05}.  Various observing programs to understand optically faint infrared sources using spectra from the $Spitzer$ Infrared Spectrometer (IRS; Houck et al. 2004) found luminous ULIRGS to redshifts z $\sim$ 3 \citep[e.g.][]{hou05,yan07,saj07,wee09}. This $Spitzer$-discovered population of high redshift ULIRGs has large infrared to optical flux ratios [$f_{\nu}$(24 \ums) $>$ 1 mJy and $R$ $>$ 24] and has been labeled ``dust obscured galaxies" (DOGS; Dey et al. 2008).  DOGs are the high redshift, most luminous examples of ULIRGs.  

The requirement for a high redshift population of AGN with substantial multiwavelength extinction was initially demonstrated by the spectrum of the X-ray background, requiring a harder spectrum than could be explained by the population of unobscured type 1 AGN and quasars \citep{set89}.  This harder spectrum was explained by a population of ``Compton-thick" AGN having sufficient column densities of gas ($\ga$ 10$^{24}$ cm$^{-2}$) to absorb soft X-rays \citep[e.g.][]{fab99,gil07}. Various studies confirmed the presence of a population of obscured AGN (usually considered as type 2 AGN) that outnumbers unobscured AGN (type 1 AGN) by a factor of a few \citep{mai95,wil00,ale03,mar06,hic07}.    

The most luminous DOGS also have multiwavelength spectral characteristics of AGN, including X-ray luminosity, "power law" near infrared spectra without indication of stellar spectral features, and mid-infrared spectra without the polycyclic aromatic hydrocarbon (PAH) features that characterise starbursts \citep[e.g.][]{brn06,don07,fio08,bus09}. A natural conclusion is that these sources are a population of AGN that represents those type 2 AGN which are most extreme in obscuration and luminosity .   

The major advantage of using infrared luminosities to characterise sources is that all AGN can be compared consistently without the uncertain extinction corrections necessary for luminosity measures at shorter wavelengths.  For example, dust extinction at 0.2 \um is a factor of $\sim$ 100 greater than at 20 \um \citep{dra03a,dra03b}.  With the completion of the $Spitzer$ mission and the full archive of spectra from the IRS, many previously defined samples of AGN and quasars can now be uniformly measured in the mid-infrared from $\sim$ 5 \um to $\sim$ 35 \um.  Thousands of low resolution spectra are available in the public archive ``The Cornell Atlas of Spitzer/IRS Spectra" (CASSIS; Lebouteiller et al. 2011\footnote{http://cassis.astro.cornell.edu/atlas; CASSIS is a product of the Infrared Science Center at Cornell University.}), a database of IRS spectra extracted in an optimal and consistent manner.  

In the present paper, we describe and calibrate an infrared AGN classification, measure infrared luminosities, and compare to virial black hole masses and optically-derived bolometric luminosities by analyzing CASSIS spectra for the following samples: 

\noindent 1. The uniform sample of AGN in \citet{tue10} discovered because of high energy X-rays in the all sky sample from the Burst Alert Telescope (BAT) on the $Swift$ Small Explorer \citep{geh04}; 125 of these have IRS low resolution spectra. 

\noindent 2. The AGN with the most accurate determinations of virial black hole masses, using the reverberation mapping technique to determine scale size of the broad line region \citep{pet04}; 32 have IRS low resolution spectra (many of these are also in the BAT sample). 

\noindent 3. The 100 type 1 quasars most luminous in the mid-infrared [$\nu L_{\nu}$(7.8 \ums)]; these are found by combining an empirical spectral template determined with the IRS for type 1 AGN and quasars with photometry at 22 \um from the Wide-Field Infrared Survey Explorer (WISE, Wright et al. 2010) for optically discovered quasars in the Sloan Digital Sky Survey (SDSS; Gunn et al. 1998, Schneider et al. 2010).  

The results give consistent measures of dust luminosity for local AGN and for the most luminous type 1 quasars to z = 5.  Results are compared to the infrared luminosities for the most luminous DOGs discovered both by $Spitzer$ and by WISE.  

\section{IR Spectral Classification and Dust Luminosities}

The mid-infrared spectra of AGN are dominated by the continuum emission from dust and show direct evidence of either dust emission or absorption, revealed primarily by the 9.7 \um silicate feature (Figure 1). The presence of silicate absorption means there is cooler dust between the observer and the hotter dust responsible for the infrared continuum.  The infrared spectra of absorbed AGN are consistent with observing an AGN behind dust clouds, with dust on the inner side of the absorbing clouds heated by the AGN \citep[e.g.][]{ima07}.  When classifiable optically, these AGN are most often type 2 \citep{hao07} but often are not classifiable optically because of strong extinction of the AGN. 

By contrast, type 1 AGN and quasars generally show silicate emission \citep[e.g.][]{hao05,hao07}, which indicates that the hot dust is observed without intervening cool dust. This arises when the circumnuclear dust clouds are observed sufficiently face on that the heated side of the clouds is directly observed so silicate emission is seen.  While orientation effects alone may not explain all the spectroscopic differences among AGN, the success of dusty torus models in explaining all AGN types indicates that circumnuclear dust is a general characteristic of AGN \citep{shi06,ra11}. 

The geometrical dust distributions required to explain silicate absorption and emission observed in IRS spectra, and how this relates to optical classifications, have been carefully discussed \citep{lev07,sir08,tho09,eli12}. The essential conclusions are that the only geometries ruled out by observations are spherically symmetric and uniform dust screens. Silicate absorption can take place through either uniform screens or clumpy dust distributions, but silicate emission cannot be observed unless some lines of sight do not pass through dust which is cooler than the emitting dust. Presence of silicate emission requires, therefore, either toroidal distributions of dust or spherical distributions of clumpy dust so that there are some lines of sight where an observer can see directly the emitting dust heated by the AGN.  

Detailed models of dust distributions including optical depth effects for absorbing clouds are not completely consistent with the simplest ``unified theory" whereby all AGN are fundamentally similar but differ in appearance from type 1 to type 2 only because of orientation to the observer.  A distribution of dust geometries is required, and the most extreme examples derive from the most extreme geometries with the largest covering factors for the dust \citep{eli12}.  For our purposes in the present paper, the most important question is how well the infrared luminosity from dust, $L_{IR}$, measures the total intrinsic luminosity, $L_{bol}$, generated by the AGN. The ratio $L_{IR}$/$L_{bol}$ is a measure of the covering factor of the dust clouds which absorb the primary radiation from $L_{bol}$ and reemit as $L_{IR}$.

Although the presence of silicate absorption or emission in infrared spectra generally correlates with type 2 and type 1 optical classification, the qualitative nature of optical classifications and the dependence of optical parameters on extinction corrections provides a motive for classification based only on the infrared silicate spectrum.  In the present paper, we quantitatively classify sources depending on the measured strength of the silicate absorption or emission.  We call sources ``obscured AGN" if they have silicate absorption and "unobscured AGN" if they have silicate emission.  This single parameter AGN classification based on silicate strength is described below.  

\subsection{AGN Classification by Silicate Strength and Mid-Infrared Luminosities}

\begin{figure}

\figurenum{1}
\includegraphics[scale= 0.8]{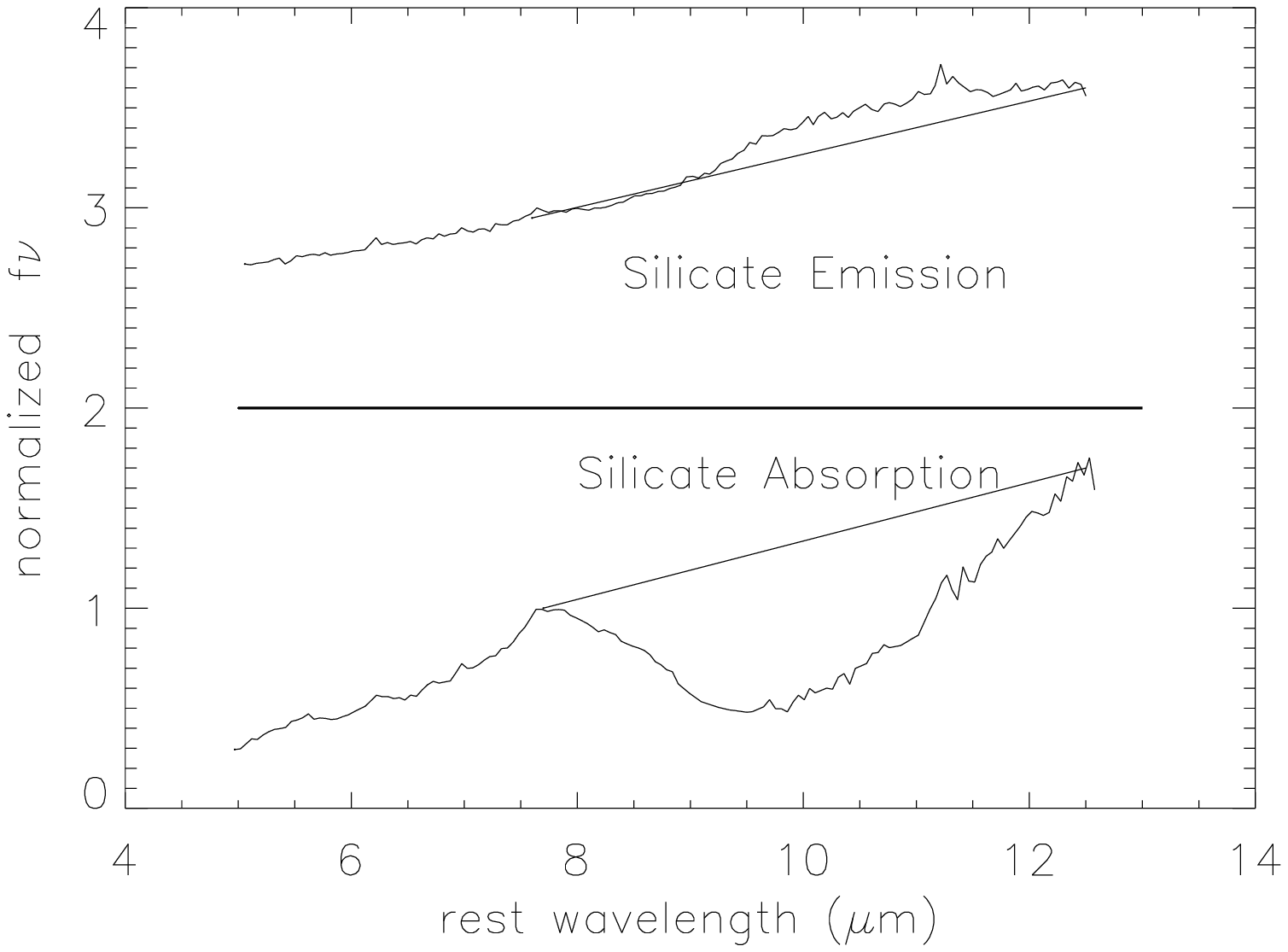}
\caption{Average of silicate emission AGN and silicate absorption AGN from the IRS spectra in \citet{sar11}, showing the 9.7 \um silicate feature.  The strength of the silicate feature is defined as the observed flux density at 10 \um compared to the extrapolated continuum between 7.8 \um and 13 \um, shown by the solid lines.   Spectra shown are normalized to $f_{\nu}$(7.8 \ums) = 1 mJy; zero levels are offset by 2 mJy so horizontal line is zero level for upper spectrum.  With this definition, the silicate strength of the upper spectrum is 1.12 and of the lower spectrum is 0.42. Mid-infrared luminosities are defined using flux $\nu f_{\nu}$(7.8 \ums) at the observed wavelength corresponding to rest frame 7.8 \um. } 

\end{figure}

The infrared classification and luminosity parameters which we utilize are illustrated in Figure 1.  The classification is a measure of the height or depth of the silicate feature measured at 10 \ums.  As illustrated in Figure 1, this measure of silicate strength assumes an extrapolated dust continuum between 7.8 \um and 13 \ums, and the silicate feature is measured relative to this continuum. The strength of the silicate feature is defined as $f_{\nu}$(10 \um observed)/$f_{\nu}$(10 \um continuum), for $f_{\nu}$(10 \um continuum) extrapolated linearly between $f_{\nu}$(7.8 \ums) and $f_{\nu}$(13 \ums), as shown. 

This parameter is designed primarily to allow estimation of silicate absorption in DOGs at z $\ga$ 2, for which IRS spectra have limited wavelength coverage and poor signal to noise (S/N).  Various alternatives for measuring silicate strength when longer, reliable continuum baselines can be used are discussed by \citet{spo07} and \citet{sir08}.  Our measurement is similar to one of the continuum fits they illustrate, except that we do not express the silicate strength as an optical depth, and we always measure at a single wavelength instead of seeking the wavelength where the feature is strongest.

The observed mid-infrared luminosity is determined as $\nu L_{\nu}$(7.8 \ums) (rest frame).  This luminosity parameter is chosen because it is the most reliable flux measurement for dusty sources, especially faint sources at z $\ga$ 2 with IRS spectra (summarized by \citet{wee09}), because this wavelength is a localized continuum maximum between absorptions on either side for heavily obscured sources (as seen in Figure 1).  This luminosity parameter is especially important for highly obscured, optically faint DOGs when optical spectra cannot be obtained.  For silicate emission sources, such as optically discovered quasars, measuring $\nu L_{\nu}$(7.8 \ums) allows an unambiguous luminosity comparison with the obscured sources. 

The greatest uncertainties in measuring $\nu L_{\nu}$(7.8 \ums) or the silicate strength in this way arise when PAH features from a starburst are present, because the flux density at 7.8 \um is not then a purely continuum measurement but also includes the peak of the 7.7 \um PAH feature.  For this reason, we do not measure silicate strength or $\nu L_{\nu}$(7.8 \ums) for any source with EW(6.2 \um PAH) $>$ 0.05 \ums, as discussed below in section 2.2. As can be seen by examining individual spectra in CASSIS, uncertainties in the measure of silicate strength are all in the direction that silicate emission would be underestimated and silicate absorption overestimated.  This is partly because any PAH contribution would artificially raise the assumed continuum at 7.8 \um and partly because the peak of silicate emission moves to wavelengths longer than 10 \um for sources with strong silicate emission.  

\subsection{Infrared Classification for the BAT AGN Sample}

The infrared AGN classification derived from silicate strength is now demonstrated for a uniform and independent AGN sample using comparisons between CASSIS infrared spectra and characteristics of the uniform sample of AGN defined by hard X-rays (14-195 keV) from the $Swift$ BAT all sky survey \citep{tue10}.   As those authors explain, this is an unbiased AGN sample discovered without regard to optical classification, using only the presence of hard X-rays.  Consequently, it is an excellent sample for objective tests of the infrared classification criteria.  Of the 234 sources listed in the BAT sample having "Sy" or "quasar" optical classifications (Table 5 of Tueller et al.), 125 have IRS spectra in CASSIS.  Results for these AGN are summarized in Table 1.

\begin{deluxetable}{lccccccccc} 
\tablecolumns{10}
\tabletypesize{\footnotesize}
\rotate
\tablewidth{0pc}
\tablecaption{Observed Properties of BAT AGN}
\tablehead{
 \colhead{No.} &\colhead{Name\tablenotemark{a}}& \colhead{AOR\tablenotemark{b}} &\colhead{flux(X)\tablenotemark{c}} & \colhead{$L$(X)\tablenotemark{d}} & \colhead{f([NeIII])\tablenotemark{e}} &\colhead{f([OIV])\tablenotemark{f}} & \colhead{EW(6.2 \ums)\tablenotemark{g}} & \colhead{$f_{\nu}$(7.8 \ums)\tablenotemark{h}} & \colhead{silicate\tablenotemark{i}}
\\
\colhead{} & \colhead{} & \colhead{} & \colhead{10$^{-11}$} & \colhead{log erg s$^{-1}$} & \colhead{10$^{-21}$} & \colhead{10$^{-21}$} & \colhead{\ums} & \colhead{mJy} & \colhead{}
}
\startdata

1 & Mrk 335 & 14448128 			& 2.47	& 43.57	 & \nodata	&\nodata& $<$0.01	&  128	& 1.08   \\   	
2 & Mrk 1501	& 4857088,14188544	& 4.10	& 44.91	 & 3.9		& 8.4	& $<$0.01	&   53	& 1.0  \\
3 & NGC 235A           &	20342016	& 	4.12	& 43.67	& 8.8		& 23.1	& 0.084 &  79	& \nodata  \\
4 & Mrk 348		& 17957376		& 13.66	& 43.84	 & 23.6		& 18.5	& $<$0.01	&  165	& 0.76  \\
5 & Mrk 1148	& 14189056		& 2.80	& 44.44	 & 0.68		& 0.35	& $<$0.01	&  10.7	& 1.16  \\	
6 & Mrk 352		& 26482944 		& 4.16	& 43.31	 & \nodata		& \nodata	& 0.03	&  6.2	& 1.36  \\	
7 & 3C 033		& 11295744		& 4.06	& 44.54	 & 3.59		& 6.39	& $<$0.01	&  24	& 0.78  \\	
8 & Fairall9        & 18505984,28720896	& 5.07	& 44.42	 & 3.9		& 5.1	& $<$0.01	& 197	& 1.11  \\
9 & NGC 526a	& 18525184,4867328	& 5.96	& 43.69	 & 10.1		& 17.4	& $<$0.01	& 98	& 0.9  \\
10 & NGC 612		& 18945280		& 5.35	& 44.04	 & 3.1		& 5.2	& 0.22	&  70	& \nodata	  \\
11 & ESO 297		& 18944768		& 7.37	& 44.03	 & 2.6		& 3.3	& 0.036	&  35	& 0.72  \\		
12 & NGC 788		& 18944512		& 9.33	& 43.59	 & 13.9		& 18	& $<$0.01	&  61	& 0.77  \\
13 & Mrk 1018	& 15076096		& 3.61	& 44.18	 & 1.6		& 1.8	& $<$0.01	&  40	& 1.19  \\
14 & IC 1816		& 26485248		& 2.58	& 43.22	 & 19.5		& 16.7	& 0.045	&  50	& 0.64 \\
15 & NGC 973		& 26485760		& 3.09	& 43.26	 & 9.0		& 20	& $<$0.01	&  56	& 0.40 \\
16 & NGC 985		& 13022720		& 3.45	& 44.17	&  13		& 22	& $<$0.01	&  100	& 0.95  \\
17 & ESO 198-024	& 26491904		& 4.98	& 44.38	 & \nodata		& \nodata	& $<$0.01	&  22	& 1.0  \\
18 & NGC 1052	& 18258688		& 3.75	& 42.32	 & 14.3		& 6.3	& $<$0.01	& 70	& 0.94  \\
19 & ESO 417-G006 	& 26486784 		& 3.06	& 43.26	 & 2.6		& 3.9	& $<$0.01	&  9.3	& 0.79  \\
20 & NGC 1275	& 3753984			& 6.85	& 43.68	 & 27		& $<$5	& $<$0.01	& 232	& 1.0   \\
21 & NGC 1365	& 9075200,8767232		& 7.19	& 42.68	 & 74		& 131	& 0.059& 	 504	& \nodata	  \\	
22 & ESO 548		& 18943744		& 4.57	& 43.33	 & 2.8		& 2.6	& $<$0.01	&  92	& 1.21  \\
23 & 2MASX J03502377	& 26496512		& 3.09	& 43.98	 & 2.0		& 1.1	& 0.49	&  18	& \nodata  \\
24 & ESO 549		& 26489600		& 2.65	& 43.62	 & 4.5		& 9.5	& 0.23	&  113	& \nodata  \\
25 & ESO 157 G23	& 26498816		& 3.02	& 44.13	 & 2.2		& 4.3	& $<$0.01	&  22.7	& 1.0  \\
26 & 3C 120		& 18505216,4847360	& 11.89	& 44.48	 & 26.9		& 91	& $<$0.01	&  138	& 1.0  \\	  
27 & MCG-02-12-050	& 26498560		& 2.11	& 43.81	 & 1.9		& 2.5	& $<$0.01	&  30	& 1.0  \\
28 & CGCG 420	& 26482176		& 4.08	& 43.91	&  5.7		& 9.9	& $<$0.01	&  171	& 1.0  \\
29 & 2MASX J05054575	& 26490368		& 7.16	& 44.31	 & 2.5		& 1.8	& $<$0.01	&  36	& 0.83  \\
30 & Ark 120         & 18941440 		& 7.08	& 44.23	 & 2.6		& 3.0	& $<$0.01	&  177	& 1.19  \\
31 & PictorA		& 4673792			& 3.78	& 44.03	 & 1.9		& $<$0.5	& $<$0.01	&  31	& 1.27  \\		
32 & IRAS 05218	& 26497536		& 2.50	& 44.15	 & 7.3		& 15.7	& $<$0.01	&  73	& 0.95  \\
33 & NGC 2110	& 4851456,18509312	& 35.01	& 43.67	 & 51		& 38	& 0.013	&  155	& 1.0  \\
34 & 2MASX J05580206 & 18943232		& 3.99	& 44.02	 & 4.7		& 4.5	& $<$0.01	&  305	& 0.79  \\
35 & ESO 005-G004	& 18947328 		& 4.48	& 42.59	 & 9.3		& $<$3.0	& 0.06	&  79	& 0.21  \\	
36 & Mrk 3		& 3753472			& 15.65	& 43.81	 & 188		& 178	& $<$0.01	&  235	& 0.63  \\
37 & ESO 426		& 26495232		& 2.72	& 43.49	 & 3.0		& 12.8	& $<$0.01	&  43	& 0.86  \\	
38 &  UGC 03601	& 26493696		& 4.38	& 43.46	 & 7.4		& 12.3	& $<$0.01	&  19 & 	0.92  \\
39 & Mrk10		& 26498304		& 3.12	& 43.79	 & 6.2		& 20.3	& $<$0.01	&  22.7	& 0.83  \\
40 & IC 486 		& 18971136		& 3.22	& 43.73	 & 5.4		& 11.3	& 0.020	&  42	& 0.74   \\ 
41 & Phoenix		& 25408256		& 5.34	& 43.34	 & 35		& 18	& $<$0.01	& 202	& 0.78  \\	
42 & 2MAS X0904	& 26495744		& 1.91	& 43.78	 & 2.1		& 3.2	& 0.117	&  13	& \nodata	 \\	
43 & 2MAS X0911	& 26496256		& 1.81	& 43.47	 & 2.9		& 6.4	& 0.070	&  25	& \nodata  \\
44 & MCG-01-24-012	& 18945024		& 4.58 	& 43.60	 & 6.2		& 10.5	& $<$0.01	&  53 & 	0.53  \\
45 & MCG+04-22-042	& 26491392		& 4.46	& 44.03	 & 5.9		& 8.2	& $<$0.01	&  70	& 1.11  \\
46 & Mrk 110		& 14189824		& 6.15	& 44.25	 & 3.0		& 4.1	& $<$0.01	&  44	& 1.17  \\		
47 & Mrk 705		& 14203392		& 2.13	& 43.62	 & 5.1		& 5.4	& 0.021	&  71	& 1.0  \\
48 & NGC 2992	& 26122240,4934144  	& 4.82	& 42.80	 & 88		& 101	& 0.085	& 130	& \nodata  \\
49 &  MCG -05-23-016	& 26484992 		& 20.77	& 43.52	 & 18.1		& 27	& $<$0.01	&  337	& 0.69  \\	
50 & NGC 3081	& 4851968,18509824	& 10.24	& 43.16	&  44		& 97	& $<$0.01	&  107	& 0.75  \\	
51 & NGC 3079	& 3755520			& 3.44	& 42.02	 & 29		& 14	& 0.41	&  864	& \nodata  \\
52 & ESO 374 G44	& 26497792		& 2.82	& 43.72	 & \nodata		& 22	& $<$0.01	&  25.7	& 0.8   \\ 
53 & NGC 3227	& 4934656			& 14.13	& 42.67	 & 64		& 65	& 0.101	&  296	& \nodata  \\
54 & NGC 3281	& 4852224			& 9.01	& 43.36	 & \nodata		& \nodata	& $<$0.01	&  481	& 0.27  \\ 
55 & Mrk 417		& 18946048		& 3.74	& 43.97	 & 3.2		& 5.2	& $<$0.01	&  32	& 0.88  \\	
56 & UGC 06527	& 26868992		& 2.68	& 43.68	&  8.7		& 10.7	& $<$0.01	&  52	& 0.93  \\
57 & NGC 3783	& 4852736,18510592	& 19.45	& 43.61	 & 25		& 34	& $<$0.01	&  317	& 1.0  \\
58 & UGC 06728	& 26483712  		& 2.95	& 42.44	 & 1.3		& 3.3	& $<$0.01	&  20	& 1.11  \\
59 & 2MASX J11454045	& 26484736		& 6.38	& 44.20	&  6.1		& 24	& $<$0.01	&  40	& 1.0  \\
60 &  NGC 3998		& 10512896		& 3.03	& 41.91	 & \nodata		& \nodata	& 0.01	&  40  	& 1.29  \\
61 & NGC 4051		& 14449152		& 4.34	& 41.71	 & \nodata		& \nodata	& 0.044	&  281	& 0.94  \\
62 & Ark 347		& 26483456		& 3.85	& 43.64	 & 12.2		& 28.5	& $<$0.01	&  50	& 1.0  \\
63 & NGC 4102	& 18941952		& 2.58	& 41.66	 & 37		& $<$26	& 0.311	& 1080	& \nodata  \\
64 & NGC 4151	& 3754496			& 62.23	& 43.18	 & 230		& 214	& $<$0.01	&  970	& 0.96  \\	
65 & NGC 4235	& 22086144		& 2.40	& 42.54	 & 6.3		& 3.7	& $<$0.01	&  23	& 1.0  \\
66 & Mrk 766 	& 14448896		& 2.42	& 42.96	 & \nodata		& \nodata	& 0.042	& 199	& 0.72  \\ 
67 & M106		& 4934912,18526208	& 2.80	& 41.14 	 & 12.0		& 6.3	& 0.024	&  79	& 1.13  \\
68 & Mrk 50		& 26496768		& 3.53	& 43.64	 & \nodata		& \nodata	& 0.01	&  17	& 1.16  \\
69 & NGC 4395	& 17792256		& 3.12	& 40.89	 & 2.5		& 4.4	& $<$0.01	&  3.7	& 1.0  \\ 
70 & NGC 4388	& 4852992,18510848 	& 34.64	& 43.74	 & 134		& 295	& 0.044	& 238	& 0.40  \\
71 & 3C 273		& 19718656 		& 38.35	& 46.42	 & 3.2		& 7.8	& $<$0.01	&  276	& 1.0  \\
72 & NGC 4507	& 18511104,4853248	& 22.51	& 43.85	 & 29		& 35	& $<$0.01	&  349	& 1.0  \\
73 & ESO 506		& 18941696		& 13.75	& 44.29	 & 4.3		& 3.9	& $<$0.01	&  131	& 0.32  \\
74 & NGC 4593	& 4853504			& 9.79	& 43.25	 & \nodata		& \nodata	& $<$0.01	&  204	& 1.04  \\ 
75 & NGC 4686	& 26492160		& 3.08	& 43.29	 & 1.7		& 1.1	& $<$0.01	& 17	& 0.77   \\
76 & SBS 1301		& 26498048		& 4.02	& 43.92	 & 0.36		& 0.28	& $<$0.01	&  10	& 1.21  \\
77 & NGC 4945		& 8768000,8769280		& 32.98	& 42.41	 & 109		& \nodata	& 0.21	& 2940	& \nodata  \\
78 & ESO 323-077	& 18942720		& 4.70	& 43.38	 & 17.2		& 29	& 0.029	& 426	& 0.96  \\
79 & NGC 4992		& 26490880		& 6.16	& 43.95	 & $<$0.8		& 1.1	& $<$0.01	&  65	& 0.34   \\
80 & MCG-03-34-064	& 20367615 		& 3.15	& 43.29	 & 121		& 106	& $<$0.01	&  410	& 0.74  \\
81 & Cen A		& 8766464,26121984	& 92.62	& 42.83	 & \nodata		& \nodata	& 0.01	&  891& 	0.39  \\	
82 & MCG-06-30-015	& 4849920			& 7.82	& 43.02	 & 8		& 18	& $<$0.01	&  125	& 0.92	 \\ 
83 & NGC 5252		& 18946304		& 8.18	& 43.99	 & 6.0		& 8.7	& $<$0.01	&  32	& 1.0  \\
84 & IC 4329A 	& 18506496		& 33.08	& 44.28	&  63		& 111	& $<$0.01	&  616	& 0.93  \\
85 & UM 614		& 26497024		& 2.24	& 43.74	 & 2.1		& 5.7	& $<$0.01	&  20	& 1.0  \\
86 & Mrk 279		& 7616512			& 5.30	& 44.05	 & 8.7		& 13.6	& 0.008	&  111	& 1.0  \\
87 & Circinus	& 9074176			& 27.48	& 42.10	 & \nodata		& \nodata	& 0.033	& 7850	& 0.28  \\
88 & NGC 5506	& 18512896,4855040	& 25.64	& 43.34	 & 157		& 240	& 0.007	& 1030	& 0.43  \\
89 & NGC 5548	& 18513152,4855296	& 8.08	& 43.73	 & 9.5		& 9.7	& 0.021	& 110	& 0.97  \\
90 & ESO 511		& 26491136		& 4.71	& 43.73	 & 1.8		& 0.8	& $<$0.01	&  40	& 1.38  \\
91 & NGC 5728	& 18945536		& 10.54	& 43.31	 & 58		& 109	& 0.12	& 115	& \nodata  \\
92 & Mrk 841		& 3761664			& 2.93	& 43.96	 & 9.7		& 20	& $<$0.01	&  92	& 0.94   \\
93 & Mrk 1392		& 15079168		& 1.99	& 43.78	 & 8.9		& 17.1	& $<$0.01	&  34	& 1.0  \\
94 & NGC 5899	& 22090240		& 2.15	& 42.54	 & 18		& 25	& $<$0.01	&  24	& 0.27  \\
95 & NGC 6240	& 4985600			& 7.30	& 44.00	 & 66		& 43	& 0.28	&  463	& \nodata  \\
96 & NGC 6300	& 22091264		& 10.35	& 42.50	 & 15		& 29	& 0.026	&  319	& 0.26  \\
97 & HB89(1821)	& 4676096			& 1.74	& 45.69	 & 11.0		& 23	& $<$0.01	&  127	& 1.0  \\
98 & 3C 380		& 4581888			& 2.56	& 46.73	 & 0.48		& \nodata	& \nodata	&  15	& 1.11  \\
99 & 3C 382		& 11306496		& 8.42	& 44.83	 & 1.6		& 1.7	& $<$0.01	&  83	& 1.11  \\
100 & Fairall49	& 18507520		& 2.93	& 43.43	 & 41		& 37	& 0.021	&  387 & 	0.72  \\
101 & ESO 103-G035	& 18505728,4847872	& 11.14	& 43.64	 & 31		& 31	& $<$0.01	&  176	& 0.39  \\        
102 & 3C 390.3		& 4673024			& 10.97	& 44.92	 & 2.8		& 1.9	& $<$0.01	&  62	& 1.0   \\	
103 & Fairall51	& 26489088		& 4.57 	& 43.31	 & 15.9		& 22.2	& 0.025	& 239	& 1.0  \\
104 & ESO 141 G55	& 26489856		& 5.27	& 44.20	&  4.3		& 6.9	& $<$0.01	& 113	& 1.18  \\
105 & NGC 6814		& 22091776		& 7.82	& 42.67	 & 15		& 18	& $<$0.01	&  69	& 1.0  \\
106 & CygnusA		& 4673280			& 12.23	& 44.96	&  42		& 63	& $<$0.01	&  54	& 0.43  \\
107 & MCG 0448		& 20349184		& 8.77	& 43.58	 & 22.5		& 17.6	& 0.48	&  347	& \nodata  \\
108 & Mrk 509		& 4850432,18508288	& 9.44	& 44.41	 & 13.7		& 26	& 0.019	&  191	& 0.97  \\
109 & IC 5063		& 18506752		& 8.59	& 43.39	 & 73		& 91	& $<$0.01	& 454	& 0.69  \\
110 & 3C 433		& 11307264		& 1.74	& 44.66	 & 2.7		& 7.0	& $<$0.01	&  47	& 0.42  \\
111 & Mrk 520		& 26490112		& 3.58	& 43.76	 & 31		& 50	& 0.10	&  147	& \nodata  \\
112 & NGC 7172	& 4856064,18513920	& 18.11	& 43.48	 & 20		& 43	& 0.036	&  275	& 0.13  \\		
113 & NGC 7213	& 18514176,4856320	& 5.75	& 42.64	 & 15		& 2.2	& $<$0.01	&  102	& 1.44   \\	
114 & NGC 7314	& 4856576,18514432	& 4.63	& 42.37	 & 24		& 50	& $<$0.01	&  45	& 0.57  \\
115 & Mrk 915		& 26495488		& 4.99	& 43.82	 & 18.8		& 39.2	& $<$0.01	&  33	& 0.91  \\
116 & 3C 452		& 11301632		& 3.78	& 44.79	 & 1.36		& 0.54	& $<$0.01	&  12.8	& 0.83   \\
117 & UGC 12282	& 26492672		& 2.49	& 43.21	 & 2.2		& 3.0	& $<$0.01	&  22	& 0.74  \\
118 & NGC 7469	& 3755008			& 6.66	& 43.60	 & 38		& 45	& 0.137	&  764	& \nodata  \\
119 & Mrk 926		& 4856832 		& 10.17	& 44.72	 & 5.3		& 10.0	& $<$0.01	&  59	& 0.96  \\
120 & NGC 7582	& 26121728,3855616	& 7.92	& 42.68	 & 118		& 168	& 0.285	&  134	& \nodata  \\
121 & NGC 7603	& 10870784  		& 4.70	& 43.97	 & 4.0		& 6.2	& 0.018	&  227	& 1.13  \\
122 & LCRS B232242.2	& 26497280		& 2.41	& 43.86 	 & \nodata		& \nodata	& 0.073	&  27 &  \nodata  \\
123 & NGC 7682	& 26494208		& 2.27	& 43.18	 & 9.3		& 12.1	& 0.08	&  7	& 1.0   \\
124 & NGC 7679	& 20350720		& 2.33	& 43.19	 & 30		& 27.5	& 0.379	&  261 & \nodata   \\	
125 & UGC 12741  	& 26494720		& 4.00	& 43.44	 & 1.3		& 2.2	& 0.08	&  14	& 0.40  \\

\enddata

\tablenotetext{a}{Source name as listed in BAT catalog \citep{tue10} where coordinates are also given, listed in order of R.A.  This list includes all objects listed with "Sy" or "quasar"  classification in BAT catalog ("Type" in their Table 5) which also have low resolution IRS spectra; objects listed as "blazar" are not included. }
\tablenotetext{b}{AOR number for $Spitzer$ IRS spectra available in CASSIS.}
\tablenotetext{c}{Hard X-ray flux (14-195 keV) from \citet{tue10} in units of 10$^{-11}$ erg s$^{-1}$ cm$^{-2}$. }
\tablenotetext{d}{Hard X-ray luminosity from \citet{tue10} in log erg s$^{-1}$.}
\tablenotetext{e}{Total flux of [NeIII] 15.55 \um emission line fit with single gaussian in low resolution CASSIS spectra in units of 10$^{-21}$W cm$^{-2}$.}
\tablenotetext{f}{Total flux of [OIV] 25.89 \um emission line fit with single gaussian in low resolution CASSIS spectra in units of 10$^{-21}$W cm$^{-2}$.}
\tablenotetext{g}{Rest frame equivalent width of 6.2 \um PAH feature fit with single gaussian on a linear continuum within rest wavelength range 5.5 \um to 6.9 \um.}
\tablenotetext{h} {Flux density $f_{\nu}$(7.8 \ums) from IRS low resolution spectra at observed wavelength corresponding to rest wavelength 7.8 \ums.}
\tablenotetext{i} {Strength of silicate feature defined as $f_{\nu}$(10 \um observed)/$f_{\nu}$(10 \um continuum), for $f_{\nu}$(10 \um continuum) extrapolated linearly between $f_{\nu}$(7.8 \ums) and $f_{\nu}$(13 \ums).  Values $>$ 1 correspond to silicate emission and values $<$ 1 to silicate absorption. No silicate strength is given for sources with EW(6.2 \ums) $>$ 0.05 \um because of possible contamination of the extrapolated continuum at 7.8 \um by PAH 7.7 \um emission.}

\end{deluxetable}

As initially demonstrated by \citet{gen98}, the strength of the PAH features measures the relative starburst/AGN components because PAH emission is associated with the starburst component.  In various previous summaries of IRS spectra using the same measures we use for the BAT AGN in Table 1, we have illustrated for large samples how the equivalent width of the 6.2 \um feature quantitatively correlates with the starburst/AGN classifications (Sargsyan et al. 2011, 2012 and references therein).  The summary result we adopted from these previous studies is that sources having EW(6.2 \ums) $<$ 0.1 \um have the majority of luminosity at 7.8 \um arising from an AGN, and our starburst/AGN classification from IRS spectra is based only on this EW criterion, which is also applied to luminous DOGs at high redshift \citep{wee09}.  The BAT sample in Table 1 confirms this generalized PAH classification criterion.  Of the 125 AGN with IRS spectra, only 12 have EW(6.2 \ums) $>$ 0.1 \um.  These exceptions are explainable as sources for which the AGN continuum luminosity is weak in the infrared compared to the luminosity from a circumnuclear starburst. 

Figure 2 shows the silicate absorption or emission strength from CASSIS spectra compared to optical AGN classification for $Swift$ BAT AGN in Table 1, using the optical classifications given in \citet{tue10}. (As explained in section 2.1, no silicate strength is given for the 20 sources with EW(6.2 \ums) $>$ 0.05 \um to avoid possible contamination of the 7.8 \um continuum by 7.7 \um PAH emission.) The overall consistency between infrared and optical classifications is clearly demonstrated in Figure 2.  As expected, the sources with silicate absorption are predominately type 2 and those with silicate emission predominately type 1.  

There are ambiguous sources and a wide range of silicate strength at a given optical classification.  This is not surprising given that the broad optical emission lines that classify type 1 can sometimes be seen via scattering from  intervening dust.  In fact, the observation of this scattering in polarized light led to the initial "unified theory" \citep{ant85}.  An important example of an ambiguous classification is the well studied ULIRG Markarian 231.  This object has silicate absorption of strength 0.5 in the infrared and extensive multiwavelength evidence of heavy absorption \citep{bok77,gal05} but is optically a type 1 AGN.  Such ambiguities are a primary motive for classification based only on the silicate strength, which provides a quantitative measure of the presence or absence of intervening dust between observer and AGN. 

For example, the dust covering factor can be estimated by the fraction of AGN seen with silicate absorption.  Within the BAT AGN sample, the fraction of sources with measured absorption (silicate strength less than 1 among those with measured strength in Table 1) is 58/105, or 55\%.  This implies that of all AGN, 55\% are observed through dust clouds.  This could arise from various configurations. A uniform dusty torus might cover 55\% of the solid angle around the AGN, or individual clouds in a spherical configuration could cover 55\%.  For our purposes, the numerical value of the covering factor is the important result because that is a measure of the fraction of $L_{bol}$ for AGN which is absorbed and reradiated as $L_{IR}$. 

\begin{figure}

\figurenum{2}
\includegraphics[scale=0.8]{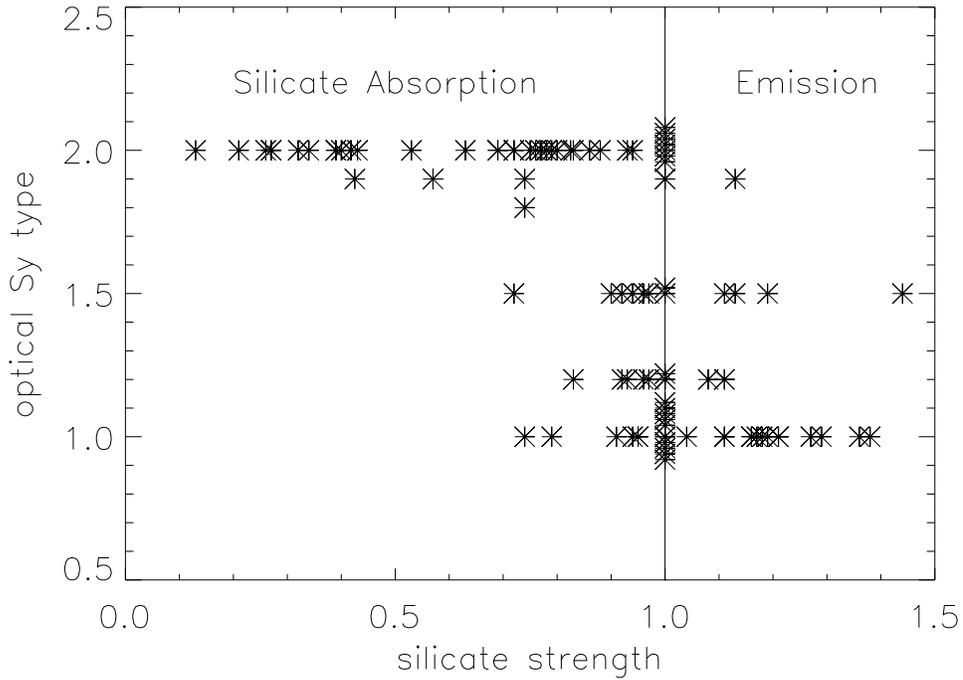}
\caption{Silicate absorption or emission compared to optical AGN classification for all $Swift$ BAT AGN having $Spitzer$ IRS spectra and optical classifications in \citet{tue10}.  Strength of the silicate feature is defined in Figure 1 and listed in Table 1. Values $>$ 1 correspond to silicate emission and values $<$ 1 to silicate absorption.  Sources having optical classes 1.0 or 2.0 and silicate strength 1.0 are artificially displaced slightly in optical class to avoid overlapping symbols.} 

\end{figure} 

\subsection{Infrared Luminosities for the BAT AGN Sample}

The primary result we desire from studying IRS spectra of the BAT sample is to calibrate how well the mid-infrared continuum luminosity $\nu L_{\nu}$(7.8 \ums) is a measure of total AGN luminosity for AGN of all classifications, because we use this mid-infrared luminosity parameter for the high redshift DOGs and other dusty quasars.   We also will compare to the emission line luminosity of the mid-infrared [OIV] 25.89 \um emission line, because of previous conclusions that this line is an indicator of intrinsic AGN luminosity \citep{rig09,wea10}.

The comparisons between infrared emission lines and BAT X-ray properties by \citet{wea10} used 79 sources with IRS high resolution spectra.  This sample is  enlarged if we use the 125 sources in Table 1 with low resolution spectra.  The [OIV] line is strong enough to measure in most of these low resolution spectra, but the lower contrast of emission lines in low resolution spectra means the measurement of line fluxes is more uncertain because of continuum noise.  To use these measurements, we need an estimate of the line flux uncertainty.  This is estimated by comparing the CASSIS low resolution measures with the high resolution measures for the objects in common.  To extend these comparisons, we also use the [NeIII] 15.55 \um line, which is of comparable strength to [OIV].  

\begin{figure}
\figurenum{3}
\includegraphics[scale=0.9]{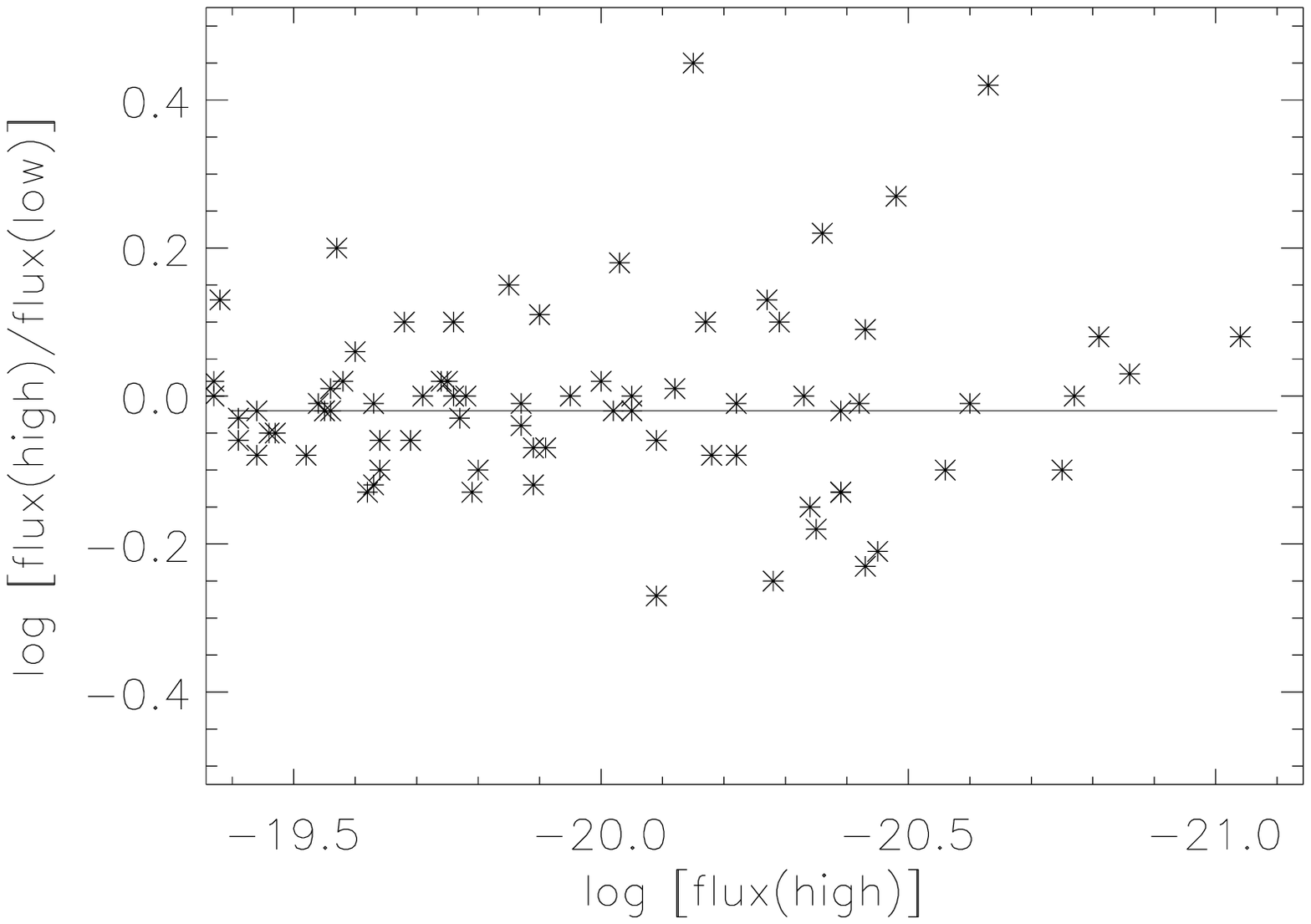}
\caption{Comparison of emission line fluxes measured in IRS low resolution CASSIS spectra of the BAT AGN compared to high resolution measures from Weaver et al. (2011) for [NeIII] and [OIV].  Units of line flux are W cm$^{-2}$. }
 
\end{figure}

\begin{figure}

\figurenum{4}
\includegraphics[scale=0.8]{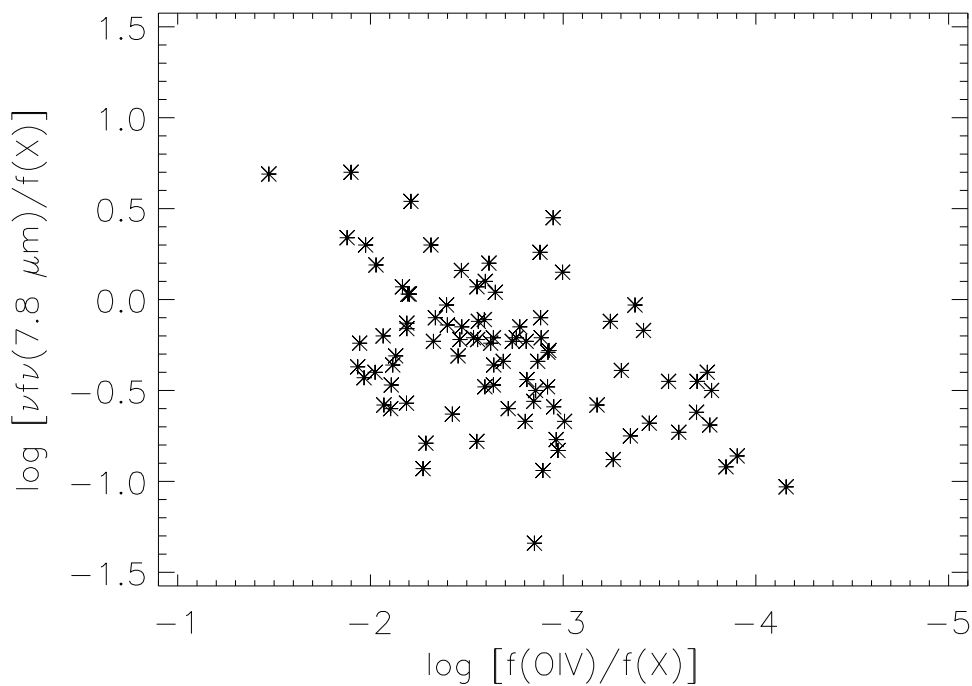}
\caption{Ratio of [OIV] 25.89 \um emission line flux to hard X-ray flux for BAT AGN, compared to ratio of mid-infrared flux to hard X-ray flux. Medians and dispersions are log [$\nu f_{\nu}$(7.8 \ums)/$f$(X)] = -0.31 $\pm$ 0.35 and log [$f$([OIV] 25.89 \ums)/$f$(X)] = -2.64 $\pm$ 0.6.  The smaller dispersion in comparing infrared with hard X-ray indicates that observed dust luminosity is a better measure of accretion luminosity than is the forbidden line luminosity.} 

\end{figure}

\begin{figure}

\figurenum{5}
\includegraphics[scale=0.8]{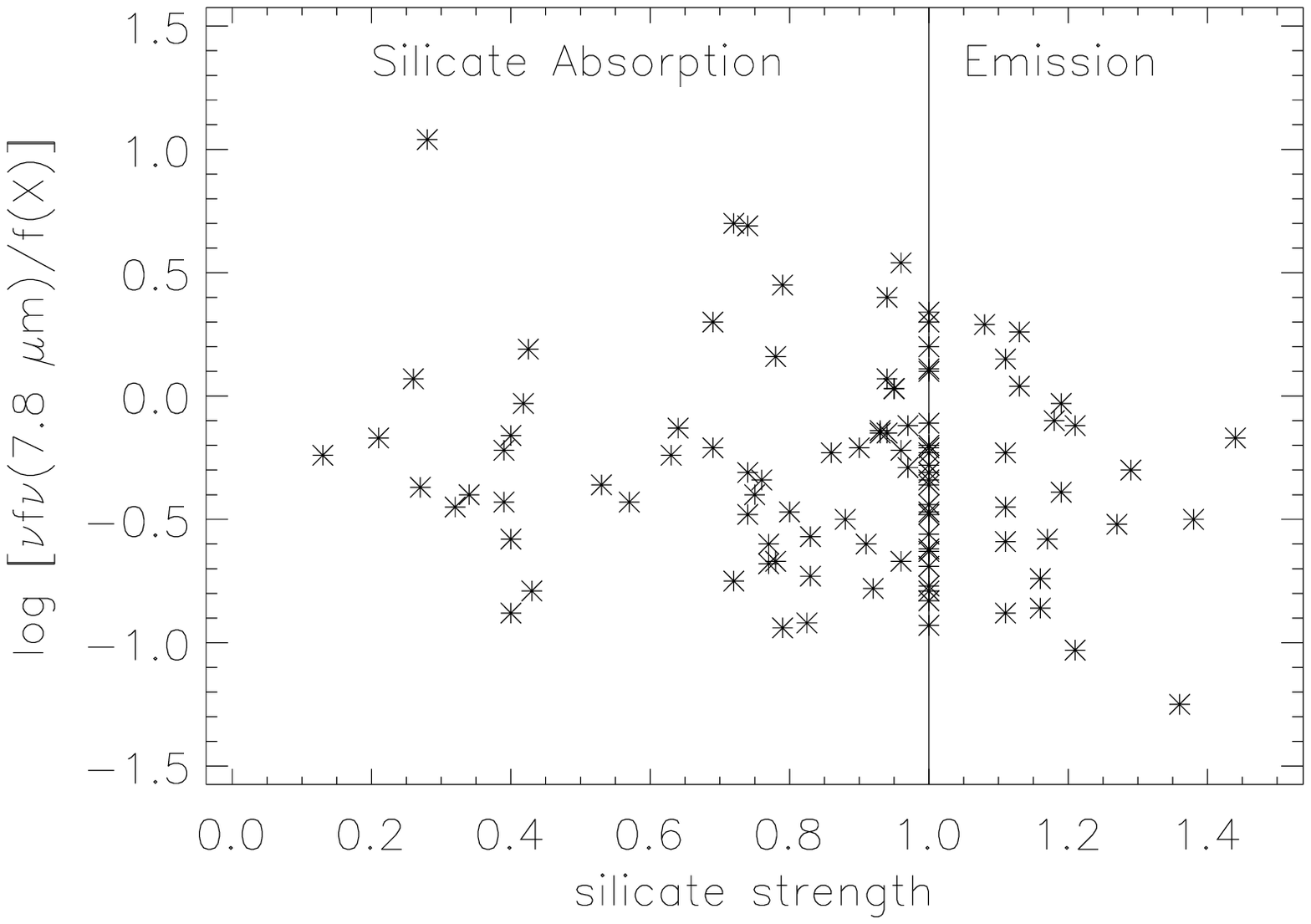}
\caption{Ratio of mid-infrared flux to hard X-ray flux for BAT AGN, compared to silicate absorption or emission strength defined in Figure 1 and listed in Table 1. Values $>$ 1 correspond to silicate emission and values $<$ 1 to silicate absorption.} 

\end{figure} 

The emission line measurements from IRS high resolution and low resolution spectra are independent.  The high resolution fitting and measurements in \citet{wea10} are based on calibrations for the high resolution spectra provided by data products from the $Spitzer$ Science Center.  The CASSIS low resolution results derive from independent calibration in the CASSIS process \citep{leb11} and from gaussian line fits done with the SMART analyis package \citep{hig04,leb09}. 

Comparisons of low resolution and high resolution line fluxes are shown in Figure 3. The median and dispersion log [flux(high)/flux(low)] = -0.02 $\pm$ 0.10.  The small systematic offset can be caused by small sample size, by differences in line fitting techniques, or by calibration differences; the total systematic difference of $\sim$ 5\% is an upper limit to calibration differences. The dispersion in the results gives the random measurement uncertainty of $\sim$ 25\%. We cannot determine from Figure 3 how much of the random uncertainty is attributed to the low resolution lines and how much to the high resolution, but we assume this dispersion to be a measure of the uncertainty only in low resolution fluxes because low resolution lines are more affected by localized continuum noise.  Therefore, the dex $\pm$ 0.10 dispersion of the ratio is taken as the uncertainty for the [OIV] line fluxes when compared to X-ray fluxes in Figure 4.  

Three measures of AGN flux given in Table 1 are compared in Figure 4: 1. Hard X-ray flux, $f$(X), from the BAT results, 2. [OIV] mid-infrared emission line flux, and 3. Mid-infrared continuum flux $\nu f_{\nu}$(7.8 \ums).  In these comparisons, flux ratios or luminosity ratios are equivalent when luminosities are compared at the same rest frame wavelengths as the observed fluxes, so we can discuss correlations using either flux ratios or luminosity ratios.  The 20 BAT AGN in Table 1 with EW(6.2 \um PAH) $>$ 0.05 \um are not used in these comparisons to avoid possible contamination of the continuum $f_{\nu}$(7.8 \ums) by the 7.7 \um PAH feature.  Medians and dispersions are log [$\nu f_{\nu}$(7.8 \ums)/$f$(X)] = -0.31 $\pm$ 0.35 and log [$f$([OIV] 25.89 \ums)/$f$(X)] = -2.64 $\pm$ 0.6. These dispersions quantitatively demonstrate that the $\nu f_{\nu}$(7.8 \ums) scales more closely with $f$(X) than does [OIV] emission, and the dispersion compared to $f$(X) provides our most unbiased calibration yet available of the precision with which the mid-infrared continuum tracks the primary luminosity from the AGN.  

The measured dispersion of $\pm$ 0.35 in dex means that the mid-infrared flux (or luminosity) predicts the hard X-ray flux (or luminosity) to a one sigma precision of about a factor of 2.  This is a cosmic dispersion illustrating differences from source to source in the various physical characteristics that affect this ratio.  Differences can arise from many effects including differences in covering factors and temperatures for the absorbing dust, and differences in the spectral energy distribution of the primary luminosity such that the luminosity at wavelengths absorbed by the dust differs in proportion to the hard X-ray luminosity.  Because of the many factors that can enter the ratio $\nu f_{\nu}$(7.8 \ums)/$f$(X) and the wide wavelength difference between these two observational measures, the resulting correlation illustrates (to us) surprising consistency among AGN.  

The next question is whether the mid-infrared continuum luminosity represents the intrinsic luminosity regardless of AGN orientation and any extinction imposed by the accident of the observer's line of sight.  This is tested in Figure 5, by considering the ratio of mid-infrared flux (or luminosity) to X-ray flux (or luminosity) for BAT AGN when compared to the strength of silicate absorption or emission.  The silicate strength is a measure of whether the AGN is observed through dust, as discussed in section 2.1.  This Figure shows no trend for the ratio log [$\nu f_{\nu}$(7.8 \ums)/$f$(X)] to depend on silicate strength; the median ratio for emission sources (strength $\ge$ 1) is -0.35 and for absorption sources (strength $<$ 1) is -0.3.  This result implies the important conclusion that the mid-infrared luminosity can be used equally well as a measure of intrinsic AGN luminosity regardless of AGN orientation and the resulting classification.

\section{Infrared Luminosities, Bolometric Luminosities, and Black Hole Masses}

The preceding sections establish the reliability of the luminosity measure $\nu L_{\nu}$(7.8 \ums) for AGN of any classification, so that AGN (or quasars) can be uniformly compared in luminosity regardless of the extinction effects that may seriously affect rest frame optical and ultraviolet measurements.  In this section, we compare this measure of infrared luminosity to other measures of bolometric luminosity for a local AGN sample and for the most infrared-luminous quasars now known.  We also compare to the most fundamental physical parameter for AGN, which is the mass of the accreting supermassive black holes.  These comparisons illustrate the differences and dispersions between very different methods of measuring luminosity.  

Long term variability studies of AGN combined with observed velocity dispersions in the broad line region (BLR) led to the "reverberation mapping" technique for measuring the virial masses of black holes in type 1 AGN \citep{pet04}. The primary set of reverberation mapped AGN contains 35 sources with black hole masses, of which 32 have CASSIS spectra.  These are summarized in Table 2 and are the local AGN sample for which we compare $\nu L_{\nu}$(7.8 \ums) to black hole masses and $L_{bol}$.

Because reverberation mapping works only if the BLR is observable, these local AGN are of necessity type 1. To extend these comparisons to the most luminous type 1 sources known, we also summarize the most infrared-luminous type 1 quasars.  These arise from the wide area SDSS survey combined with infrared photometry from WISE.  Finally, the most luminous type 1 quasars are compared to the infrared luminosities of the heavily obscured quasars among the DOGs discovered spectroscopically by $Spitzer$ and among those estimated photometrically from WISE.

\begin{deluxetable}{lccccccccc} 
\tablecolumns{10}
\tabletypesize{\footnotesize}
\rotate
\tablewidth{0pc}
\tablecaption{Observed Properties of AGN with Black Hole Masses }
\tablehead{
 \colhead{No.} &\colhead{Name\tablenotemark{a}}& \colhead{AOR\tablenotemark{b}} & \colhead{z} & \colhead{EW(6.2 \ums)\tablenotemark{c}} & \colhead{$f_{\nu}$(7.8 \ums)\tablenotemark{d}} &\colhead{$\nu L_{\nu}$(7.8 \ums)\tablenotemark{e}}& \colhead{$L_{IR}$\tablenotemark{f}}& \colhead{$L_{bol}$\tablenotemark{g}}& \colhead{BHM\tablenotemark{h}}
\\
\colhead{} & \colhead{} & \colhead{} & \colhead{} & \colhead{\ums} & \colhead{mJy}  & \colhead{log erg s$^{-1}$} & \colhead{log \ldot} & \colhead{log \ldot}& \colhead{log M$_{\odot}$}   
}

\startdata

1 & Mrk 335 & 14448128 & 0.0258 & $<$ 0.01 & 128.0 & 43.82 & 10.74 & 11.15 & 7.15 \\
2 & PG 0026+129 & 10449408,14188800 & 0.1420 & $<$ 0.01 & 14.9 & 44.39 & 11.31 & 12.33 & 8.59 \\
3 & PG 0052+251 & 4675072 & 0.1550 & $<$ 0.01 & 27.5 & 44.73 & 11.65 & 12.15 & 8.57 \\
4 & Fairall9 & 18505984,28720896 & 0.0470 & $<$ 0.01 & 197.0 & 44.52 & 11.44 & 11.32 & 8.41 \\
5 & Mrk 590 & 4850688,18508544 & 0.0264 & $<$ 0.01 & 39.0 & 43.32 & 10.24 & 10.82 & 7.68 \\
6 & 3C 120 & 18505216,4847360 & 0.0330 & $<$ 0.01 & 138.0 & 44.06 & 10.98 & 11.47 & 7.74 \\
7 & Ark 120 & 18941440 & 0.0323 & $<$ 0.01 & 177.0 & 44.15 & 11.07 & 11.24 & 8.18 \\
8 & Mrk 79 & \nodata & \nodata &  \nodata & \nodata & \nodata & \nodata & 11.03 & 7.72 \\
9 & PG 0804+761 & 9074944 & 0.1000 & $<$ 0.01 & 100.0 & 44.90 & 11.82 & 12.26 & 8.84 \\
10 & PG 0844+349 & 10449664,14189568 & 0.0640 & $<$ 0.01 & 32.0 & 44.01 & 10.93 & 11.56 & 7.97 \\
11 & Mrk 110 & 14189824 & 0.0353 & $<$ 0.01 & 44.0 & 43.62 & 10.54 & 11.01 & 7.40 \\
12 & PG 0953+414 & 4675328 & 0.2341 & $<$ 0.01 & 22.0 & 45.00 & 11.92 & 12.53 & 8.44 \\
13 & NGC 3227 & 4934656 & 0.0039 &  0.10 & 296.0 & 42.74 & 9.66 & 10.00 & 7.63 \\
14 & NGC 3516 & \nodata & \nodata &  \nodata & \nodata & \nodata & \nodata & 10.00 & 7.63 \\
15 & NGC 3783 & 4852736,18510592 & 0.0097 & $<$ 0.01 & 317.0 & 43.45 & 10.37 & 10.39 & 7.47 \\
16 & NGC 4051 & 14449152 & 0.0023 &  0.04 & 281.0 & 42.30 & 9.22 & 9.26 & 6.28 \\
17 & NGC 4151 & 3754496 & 0.0033 & $<$ 0.01 & 970.0 & 43.10 & 10.02 & 9.30 & 7.12 \\
18 & PG 1211+143 & 3760896 & 0.0809 & $<$ 0.01 & 94.0 & 44.68 & 11.60 & 12.08 & 8.16 \\
19 & PG 1226+023(3C 273) & 19718656 & 0.1583 & $<$ 0.01 & 276.0 & 45.75 & 12.67 & 13.31 & 8.95 \\
20 & PG 1229+204 & 10455296,14194944 & 0.0630 & $<$ 0.01 & 30.0 & 43.97 & 10.89 & 11.03 & 7.86 \\
21 & NGC 4593 & 4853504 & 0.0090 & $<$ 0.01 & 204.0 & 43.20 & 10.12 & 10.23 & 6.73 \\
22 & PG 1307+085 & 4735488 & 0.1550 & $<$ 0.01 & 22.0 & 44.63 & 11.55 & 12.20 & 8.64 \\
23 & IC 4329A & 18506496 & 0.0161 & $<$ 0.01 & 616.0 & 44.08 & 11.00 & 10.26 & 7.00 \\
24 & Mrk 279 & 7616512 & 0.0305 & $<$ 0.01 & 111.0 & 43.89 & 10.81 & 11.04 & 7.54 \\
25 & PG 1411+442 & 10451456,10949888 & 0.0896 & $<$ 0.01 & 75.0 & 44.68 & 11.60 & 11.90 & 8.65 \\
26 & NGC 5548 & 18513152,4855296 & 0.0172 &  0.02 & 110.0 & 43.39 & 10.31 & 10.67 & 7.83 \\
27 & PG 1426+015 & 10451712,14198272 & 0.0865 & $<$ 0.01 & 71.0 & 44.62 & 11.54 & 11.97 & 9.11 \\
28 & Mrk 817 & \nodata & \nodata &  \nodata & \nodata & \nodata & \nodata & 11.04 & 7.69 \\
29 & PG 1613+658 & 10452480,14201344 & 0.1290 & $<$ 0.01 & 74.0 & 45.00 & 11.92 & 12.11 & 8.45 \\
30 & PG 1617+175 & 10452736,14201600 & 0.1124 & $<$ 0.01 & 28.0 & 44.45 & 11.37 & 11.74 & 8.77 \\
31 & PG 1700+518 & 4675840 & 0.2920 & $<$ 0.01 & 69.0 & 45.70 & 12.62 & 12.94 & 8.89 \\
32 & 3C 390.3 & 4673024 & 0.0561 & $<$ 0.01 & 62.0 & 44.18 & 11.10 & 11.02 & 8.46 \\
33 & Mrk 509 & 4850432,18508288 & 0.0344 &  0.02 & 191.0 & 44.24 & 11.16 & 11.54 & 8.16 \\
34 & PG 2130+099 & 3761408 & 0.0630 & $<$ 0.01 & 117.0 & 44.56 & 11.48 & 11.78 & 8.66 \\
35 & NGC 7469 & 3755008 & 0.0163 &  0.14 & 764.0 & 44.19 & 11.11 & 10.68 & 7.09 \\

\enddata

\tablenotetext{a}{Source name as listed in \citet{pet04}, listed in order of R.A. Many of these sources are also in the BAT sample of Table 1.}
\tablenotetext{b}{AOR number for $Spitzer$ IRS spectra available in CASSIS.}
\tablenotetext{c}{Rest frame equivalent width of 6.2 \um PAH feature fit with single gaussian on a linear continuum within rest wavelength range 5.5 \um to 6.9 \um.}
\tablenotetext{d}{Flux density $f_{\nu}$(7.8 \ums) from IRS low resolution spectra at observed wavelength corresponding to rest wavelength 7.8 \ums.}
\tablenotetext{e}{Rest frame luminosity $\nu L_{\nu}$(7.8 \ums) in erg s$^{-1}$ determined as $\nu L_{\nu}$(7.8 \ums) =  4$\pi$D$_{L}$$^{2}$[$\nu$/(1+z)]$f_{\nu}$(7.8 \ums), for $\nu$ corresponding to 7.8 \ums, taking luminosity distances from \citet{wri06}:  http://www.astro.ucla.edu/~wright/CosmoCalc.html, for H$_0$ = 74 \kmsMpc, $\Omega_{M}$=0.27 and $\Omega_{\Lambda}$=0.73. }
\tablenotetext{f}{Total infrared luminosity $L_{IR}$ determined using calibration log [$L_{IR}$/$\nu L_{\nu}$(7.8 \ums)] = 0.51 from \citet{sar11} (Log [$\nu L_{\nu}$(\ldot)] = log [$\nu L_{\nu}$(erg s$^{-1}$)] - 33.59.)} 
\tablenotetext{g} {Optically-determined bolometric luminosity $L_{bol}$ from $\lambda L_{\lambda}$(0.51 \ums) of the AGN component in \citet{ben09}, assuming a scaling of $L_{bol}$ = 9.26$\lambda L_{\lambda}$(0.51 \ums) from \citet{ric06}.}
\tablenotetext{h}{Virial black hole mass from \citet{pet04}.} 

\end{deluxetable}

The sample of SDSS/WISE quasars was selected using the following steps.  Starting with the 105783 quasars in version 7 of the SDSS quasar catalog \citep{sch10}, all sources with z $>$ 1.5 were chosen.  The lower redshift limit was adopted to match the approximate lower redshift limit for the heavily absorbed DOGs discovered with $Spitzer$, for future comparison. (This limit is set because the DOG candidates are found when the 7.8 \um peak for absorbed quasars is within the bandpass for $Spitzer$ surveys at 24 \ums.) The upper redshift limit is the maximum z $\sim$ 5 within the SDSS. The redshift criterion results in 52761 quasars from the SDSS survey.  Of these 52761, we select the most luminous 100 as measured by $\nu L_{\nu}$(7.8 \ums).

\begin{figure}

\figurenum{6}
\includegraphics[scale=0.8]{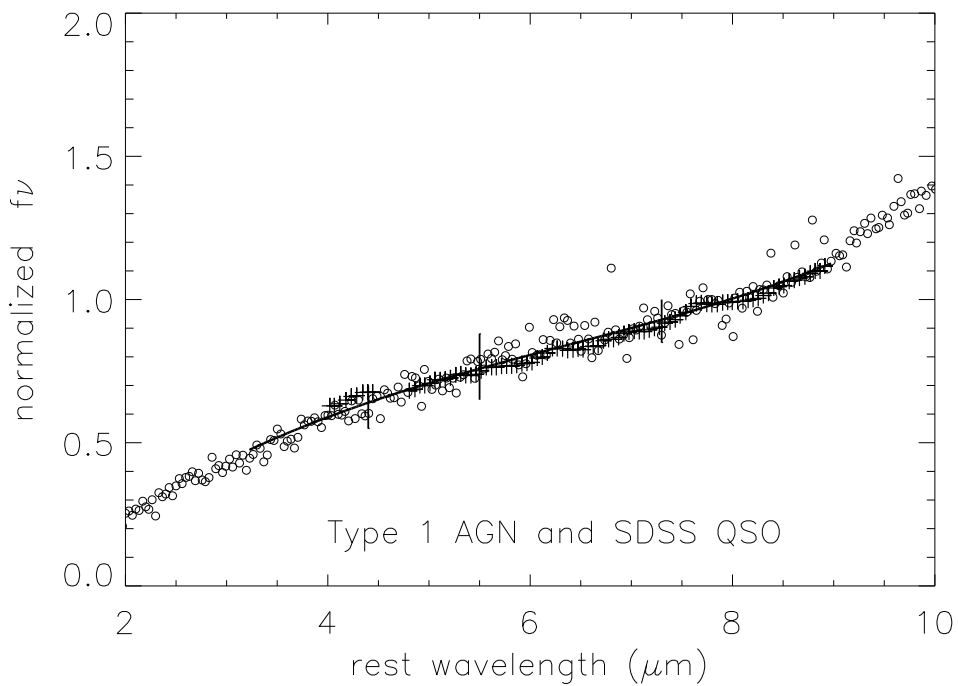}
\caption{Solid line is template used for transforming $f_{\nu}$(observed 22 \ums) to $f_{\nu}$(rest frame 7.8 \ums) for SDSS quasars with WISE photometry at 22 \ums.  Circles show median spectrum of 12 SDSS quasars also observed by IRS; crosses show median spectrum of the 33 AGN with silicate emission in Sargsyan et al. (2011).  Spectra are normalized to 1 mJy at 7.8 \ums.  Vertical error bars show dispersion among overlaid spectra for all of these type 1 AGN and quasars at wavelengths corresponding to observed frame 22 \um for z = 2, 3, and 4.} 

\end{figure}

This selection is made using $f_{\nu}$(22 \ums) from the WISE Source Catalog\footnote{http://wise2.ipac.caltech.edu/docs/release/allsky/}. In this catalog, 48651 of the 52761 SDSS quasars have WISE detections listed at some WISE wavelength, assuming a source identification to be correct if the WISE coordinate and SDSS coordinate agree to within 3~$\arcsec$.  For our uses, the necessary WISE photometry is at 22 \ums, for which we assume a real detection if $f_{\nu}$(22 \ums) $>$ 3 mJy (3$\sigma$).  Using this criterion, 16242 of the SDSS quasars are measured at 22 \ums. 

Within the redshift range 1.5 $<$ z $<$ 5, the $f_{\nu}$(22 \ums) from WISE corresponds to rest frame wavelengths 3.7 \um $<$ $\lambda$ $<$ 8.8 \ums.  The observed $f_{\nu}$(22 \ums) is transformed to $f_{\nu}$(7.8 \ums) using the empirical spectral template shown in Figure 6 to scale $f_{\nu}$ between 7.8 \um and the rest wavelength corresponding to 22 \um observed wavelength.  Among the SDSS quasars in the total sample which we use, 12 have also been observed with the IRS\footnote{SDSS 164016.09+412101.1, 163425.11+404152.5, 161007.11+535814.1, 160950.71+532909.5, 161238.26+532255.0, 081200.49+402814.3, 140323.39-000606.9, 155855.18+332318.6, 084538.66+342043.6, 030449.85-000813.4, 025905.63+001121.9, and 141546.24+112943.4; spectra are available in CASSIS.}; several of these are discussed in \citet{deo11}.  These 12 SDSS type 1 quasars with z $>$ 1.5 are combined with the infrared spectra of 33 AGN with silicate emission from \citet{sar11} to produce the template in Figure 6.  

Table 3 contains the measurements for this subsample of the 100 most luminous quasars from SDSS and WISE, as determined from $\nu L_{\nu}$(7.8 \ums). (Three of these are known gravitationally lensed quasars \citep{ina10}; these are included in the Table but not in the plots.)

Virial black hole mass estimates and $L_{bol}$ are also provided for these SDSS quasars \citep{she11}, derived using the velocity dispersions for the CIV emission line and assuming from \citet{ric06} that $L_{bol}$ = 5.15$\lambda L_{\lambda}$(0.30 \ums) for 0.7 $<$ z $<$ 1.9 and $L_{bol}$ = 3.81$\lambda L_{\lambda}$(0.135 \ums) for z $>$ 1.9.

\begin{deluxetable}{lcccccccc} 
\tablecolumns{9}
\tabletypesize{\footnotesize}

\tablewidth{0pc}
\tablecaption{Infrared Luminosities for Most Luminous SDSS Quasars}
\tablehead{
 \colhead{Source} &\colhead{Name}&\colhead{z\tablenotemark{a}} & \colhead{$f_{\nu}$(22 \ums)\tablenotemark{b}}& \colhead{$f_{\nu}$(7.8 \ums)\tablenotemark{c}}  &\colhead{$\nu L_{\nu}$(7.8 \ums)\tablenotemark{d}}&\colhead{$L_{IR}$\tablenotemark{e}}&\colhead{$L_{bol}$\tablenotemark{f}}&\colhead{BHM\tablenotemark{g}}
\\
\colhead{} & \colhead{} & \colhead{} & \colhead{mJy} & \colhead{mJy}  & \colhead{log erg s$^{-1}$} & \colhead{log \ldot} & \colhead{log \ldot}& \colhead{log M$_{\odot}$}   
}
\startdata

1 & SDSS 000654.10-001533.4 & 1.724 & 227.95 & 223.34 & 47.77 & 14.69 & 13.40 & 9.55 \\
2 & SDSS 000743.78-002432.4 & 1.540 & 104.10 & 95.22 & 47.31 & 14.23 & 13.31 & 9.27 \\
3 & SDSS 000917.68-001344.8 & 1.533 & 57.42 & 52.39 & 47.04 & 13.96 & 12.45 & 9.11 \\
4 & SDSS 001022.14-003701.2 & 3.152 & 42.72 & 57.39 & 47.64 & 14.56 & 13.77 & 9.56 \\
5 & SDSS 002614.69+143105.2 & 3.973 & 8.28 & 12.20 & 47.13 & 14.05 & 13.49 & 9.74 \\
6 & SDSS 004527.68+143816.1 & 1.992 & 34.50 & 36.76 & 47.10 & 14.02 & 13.85 & 9.42 \\
7 & SDSS 012403.77+004432.6 & 3.834 & 6.67 & 9.71 & 47.00 & 13.92 & 14.11 & 10.15 \\
8 & SDSS 012530.85-102739.8 & 3.352 & 8.94 & 12.33 & 47.01 & 13.93 & 13.98 & 9.69 \\
9 & SDSS 020950.71-000506.4 & 2.828 & 14.84 & 18.98 & 47.08 & 14.00 & 14.32 & 10.18 \\
10 & SDSS 021646.94-092107.2 & 3.716 & 7.41 & 10.65 & 47.02 & 13.94 & 13.98 & 10.27 \\
11 & SDSS 024230.65-000029.7 & 2.506 & 24.65 & 29.71 & 47.18 & 14.10 & 12.86 & 9.30 \\
12 & SDSS 041420.90+060914.2 & 2.632 & 13.55 & 16.74 & 46.97 & 13.89 & 13.62 & 10.28 \\
13 & SDSS 073502.30+265911.5 & 1.973 & 27.56 & 29.20 & 46.99 & 13.91 & 14.25 & 10.48 \\
14 & SDSS 074521.78+473436.1 & 3.220 & 10.50 & 14.23 & 47.05 & 13.97 & 14.70 & 9.94 \\
15 & SDSS 074711.14+273903.3 & 4.154 & 5.64 & 8.45 & 47.00 & 13.92 & 14.16 & 10.46 \\
16 & SDSS 080117.79+521034.5 & 3.236 & 11.83 & 16.08 & 47.10 & 14.02 & 14.45 & 10.59 \\
17 & SDSS 081331.28+254503.0\tablenotemark{h} & 1.510 & 91.93 & 83.07 & 47.23 & 14.15 & 14.33 & 9.86 \\
18 & SDSS 081855.77+095848.0 & 3.674 & 6.83 & 9.78 & 46.98 & 13.90 & 14.07 & 9.98 \\
19 & SDSS 084631.52+241108.3 & 4.743 & 4.45 & 6.97 & 47.00 & 13.92 & 13.62 & 10.03 \\
20 & SDSS 090033.50+421547.0 & 3.290 & 9.34 & 12.78 & 47.01 & 13.93 & 14.51 & 9.82 \\
21 & SDSS 090334.94+502819.3\tablenotemark{h} & 3.584 & 9.02 & 12.79 & 47.08 & 14.00 & 13.43 & 8.74 \\
22 & SDSS 090423.37+130920.7 & 2.976 & 23.54 & 30.82 & 47.32 & 14.24 & 14.18 & 9.97 \\
23 & SDSS 092819.29+534024.1 & 4.390 & 4.60 & 7.03 & 46.96 & 13.88 & 13.56 & 10.05 \\
24 & SDSS 094140.17+325703.2 & 3.452 & 7.80 & 10.88 & 46.98 & 13.90 & 13.75 & 10.15 \\
25 & SDSS 094734.19+142116.9 & 3.030 & 9.80 & 12.94 & 46.96 & 13.88 & 14.22 & 10.24 \\
26 & SDSS 095031.63+432908.4 & 1.771 & 37.59 & 37.41 & 47.01 & 13.93 & 14.10 & 9.75 \\
27 & SDSS 095841.21+282729.5 & 3.382 & 15.74 & 21.79 & 47.27 & 14.19 & 14.09 & 10.53 \\
28 & SDSS 095937.11+131215.4 & 4.056 & 6.29 & 9.34 & 47.03 & 13.95 & 14.40 & 9.72 \\
29 & SDSS 101336.37+561536.3 & 3.633 & 7.12 & 10.15 & 46.99 & 13.91 & 13.90 & 9.43 \\
30 & SDSS 101447.18+430030.1 & 3.126 & 9.06 & 12.12 & 46.96 & 13.88 & 14.57 & 10.45 \\
31 & SDSS 101549.00+002020.0 & 4.403 & 4.82 & 7.37 & 46.98 & 13.90 & 13.76 & 9.67 \\
32 & SDSS 102040.61+092254.2 & 3.643 & 7.67 & 10.95 & 47.02 & 13.94 & 13.99 & 10.30 \\
33 & SDSS 102541.78+245424.2 & 2.384 & 20.92 & 24.59 & 47.06 & 13.98 & 12.60 & 9.00 \\
34 & SDSS 102632.97+032950.6 & 3.885 & 6.75 & 9.87 & 47.02 & 13.94 & 13.10 & 9.13 \\
35 & SDSS 102714.77+354317.4 & 3.109 & 22.01 & 29.39 & 47.34 & 14.26 & 14.44 & 9.91 \\
36 & SDSS 104846.63+440710.8 & 4.347 & 4.85 & 7.38 & 46.97 & 13.89 & 13.50 & 9.58 \\
37 & SDSS 105122.46+310749.3 & 4.253 & 4.92 & 7.43 & 46.96 & 13.88 & 13.97 & 9.91 \\
38 & SDSS 105756.25+455553.0 & 4.138 & 5.29 & 7.92 & 46.97 & 13.89 & 14.33 & 10.12 \\
39 & SDSS 110352.74+100403.1 & 3.606 & 7.44 & 10.57 & 47.00 & 13.92 & 13.65 & 9.96 \\
40 & SDSS 110607.47-173113.5 & 2.572 & 18.29 & 22.34 & 47.08 & 14.00 & 14.02 & 10.25 \\
41 & SDSS 110610.72+640009.6 & 2.203 & 24.16 & 27.22 & 47.05 & 13.97 & 14.41 & 10.28 \\
42 & SDSS 111017.13+193012.5 & 2.497 & 17.58 & 21.15 & 47.03 & 13.95 & 12.24 & 8.07 \\
43 & SDSS 111038.63+483115.6 & 2.955 & 17.71 & 23.12 & 47.19 & 14.11 & 14.40 & 10.25 \\
44 & SDSS 111055.21+430510.0 & 3.822 & 9.45 & 13.73 & 47.15 & 14.07 & 13.89 & 10.26 \\
45 & SDSS 111119.10+133603.9 & 3.481 & 9.49 & 13.30 & 47.07 & 13.99 & 14.35 & 10.34 \\
46 & SDSS 112258.77+164540.3 & 3.031 & 15.06 & 19.88 & 47.15 & 14.07 & 13.76 & 10.03 \\
47 & SDSS 113017.37+073212.9 & 2.659 & 23.98 & 29.77 & 47.23 & 14.15 & 14.16 & 9.95 \\
48 & SDSS 114117.44+060332.7 & 3.294 & 9.48 & 12.98 & 47.02 & 13.94 & 12.93 & 8.97 \\
49 & SDSS 115421.69+025414.0 & 1.671 & 48.73 & 46.86 & 47.06 & 13.98 & 13.01 & 9.34 \\
50 & SDSS 115747.99+272459.6 & 2.212 & 27.38 & 30.92 & 47.10 & 14.02 & 13.12 & 9.33 \\
51 & SDSS 115906.52+133737.7 & 3.984 & 7.09 & 10.47 & 47.06 & 13.98 & 14.29 & 10.66 \\
52 & SDSS 120006.25+312630.8 & 2.989 & 11.42 & 14.99 & 47.01 & 13.93 & 14.56 & 10.23 \\
53 & SDSS 120144.36+011611.6 & 3.233 & 11.44 & 15.53 & 47.09 & 14.01 & 14.19 & 10.30 \\
54 & SDSS 120147.90+120630.2 & 3.510 & 8.71 & 12.25 & 47.04 & 13.96 & 14.32 & 9.79 \\
55 & SDSS 120447.15+330938.7 & 3.616 & 8.56 & 12.17 & 47.06 & 13.98 & 13.36 & 9.54 \\
56 & SDSS 121027.62+174108.9 & 3.610 & 9.97 & 14.17 & 47.13 & 14.05 & 13.80 & 10.67 \\
57 & SDSS 121537.88+022753.3 & 3.634 & 7.46 & 10.63 & 47.01 & 13.93 & 13.12 & 9.55 \\
58 & SDSS 121549.81-003432.1 & 2.679 & 18.06 & 22.50 & 47.11 & 14.03 & 13.88 & 9.76 \\
59 & SDSS 121930.77+494052.2 & 2.699 & 12.75 & 15.94 & 46.97 & 13.89 & 14.31 & 10.24 \\
60 & SDSS 122016.87+112628.1 & 1.881 & 47.89 & 49.38 & 47.18 & 14.10 & 13.70 & 8.63 \\
61 & SDSS 123641.45+655442.1 & 3.387 & 8.55 & 11.84 & 47.00 & 13.92 & 14.37 & 10.36 \\
62 & SDSS 123714.60+064759.5 & 2.781 & 11.75 & 14.90 & 46.96 & 13.88 & 13.53 & 9.81 \\
63 & SDSS 124551.44+010505.0 & 2.809 & 11.48 & 14.63 & 46.96 & 13.88 & 13.72 & 9.27 \\
64 & SDSS 124957.23-015928.8 & 3.638 & 9.12 & 13.00 & 47.09 & 14.01 & 14.19 & 10.40 \\
65 & SDSS 125005.72+263107.5 & 2.048 & 30.98 & 33.53 & 47.08 & 14.00 & 14.58 & 9.80 \\
66 & SDSS 125050.88+204658.7 & 3.570 & 7.35 & 10.40 & 46.98 & 13.90 & 13.13 & 10.23 \\
67 & SDSS 130502.28+052151.1 & 4.086 & 9.47 & 14.10 & 47.21 & 14.13 & 13.87 & 10.43 \\
68 & SDSS 131011.60+460124.4 & 2.134 & 22.15 & 24.53 & 46.98 & 13.90 & 14.25 & 10.06 \\
69 & SDSS 132654.96-000530.1 & 3.306 & 10.95 & 15.02 & 47.09 & 14.01 & 12.85 & 8.05 \\
70 & SDSS 132827.06+581836.8 & 3.139 & 13.95 & 18.71 & 47.15 & 14.07 & 13.53 & 8.60 \\
71 & SDSS 133335.78+164903.9 & 2.089 & 23.37 & 25.57 & 46.98 & 13.90 & 14.17 & 9.79 \\
72 & SDSS 141546.24+112943.4\tablenotemark{h} & 2.560 & 58.16 & 70.87 & 47.58 & 14.50 & 14.28 & 9.33 \\
73 & SDSS 142123.97+463318.0 & 3.363 & 8.56 & 11.82 & 47.00 & 13.92 & 14.25 & 10.40 \\
74 & SDSS 142243.02+441721.2 & 3.545 & 14.99 & 21.16 & 47.29 & 14.21 & 14.00 & 10.45 \\
75 & SDSS 142656.18+602550.8 & 3.192 & 22.77 & 30.76 & 47.37 & 14.29 & 14.69 & 10.41 \\
76 & SDSS 143352.21+022713.9 & 4.721 & 5.11 & 7.99 & 47.06 & 13.98 & 14.16 & 10.80 \\
77 & SDSS 143835.95+431459.2 & 4.611 & 7.47 & 11.60 & 47.21 & 14.13 & 14.41 & 10.38 \\
78 & SDSS 144105.53+045454.9 & 2.064 & 26.24 & 28.53 & 47.02 & 13.94 & 13.91 & 10.19 \\
79 & SDSS 144709.24+103824.5 & 3.675 & 7.38 & 10.57 & 47.01 & 13.93 & 12.88 & \nodata \\
80 & SDSS 145125.31+144136.0 & 3.102 & 13.21 & 17.63 & 47.11 & 14.03 & 13.60 & 9.68 \\
81 & SDSS 150654.55+522004.7 & 4.068 & 6.94 & 10.32 & 47.07 & 13.99 & 13.47 & \nodata \\
82 & SDSS 151352.52+085555.7 & 2.904 & 25.58 & 33.12 & 47.34 & 14.26 & 14.01 & 10.35 \\
83 & SDSS 152156.48+520238.5 & 2.208 & 21.55 & 24.31 & 47.00 & 13.92 & 14.60 & 10.11 \\
84 & SDSS 153830.55+085517.0 & 3.551 & 7.96 & 11.25 & 47.01 & 13.93 & 14.46 & 10.02 \\
85 & SDSS 154446.34+412035.7 & 3.548 & 7.43 & 10.49 & 46.98 & 13.90 & 13.29 & 10.25 \\
86 & SDSS 154938.72+124509.1 & 2.387 & 20.85 & 24.51 & 47.06 & 13.98 & 13.59 & 9.29 \\
87 & SDSS 155434.17+110950.6 & 2.936 & 18.12 & 23.58 & 47.20 & 14.12 & 12.32 & 8.56 \\
88 & SDSS 155514.85+100351.3 & 3.502 & 7.40 & 10.40 & 46.97 & 13.89 & 13.62 & 10.08 \\
89 & SDSS 155912.34+482819.9 & 3.423 & 7.79 & 10.84 & 46.97 & 13.89 & 14.15 & 10.30 \\
90 & SDSS 155952.67+192310.4 & 3.951 & 6.19 & 9.11 & 47.00 & 13.92 & 13.42 & 9.46 \\
91 & SDSS 162116.92-004250.8 & 3.703 & 7.20 & 10.33 & 47.01 & 13.93 & 14.36 & 9.57 \\
92 & SDSS 163300.13+362904.8 & 3.576 & 11.68 & 16.54 & 47.19 & 14.11 & 13.92 & 9.90 \\
93 & SDSS 163515.49+380804.4 & 1.813 & 34.79 & 35.11 & 47.00 & 13.92 & 13.78 & 9.53 \\
94 & SDSS 163909.10+282447.1 & 3.819 & 20.41 & 29.66 & 47.49 & 14.41 & 14.36 & 10.47 \\
95 & SDSS 165053.78+250755.4 & 3.341 & 14.41 & 19.85 & 47.22 & 14.14 & 13.71 & 9.18 \\
96 & SDSS 170100.60+641209.3 & 2.735 & 20.94 & 26.36 & 47.19 & 14.11 & 14.67 & 10.36 \\
97 & SDSS 212329.46-005052.9 & 2.262 & 20.37 & 23.29 & 47.00 & 13.92 & 14.36 & 10.31 \\
98 & SDSS 223808.07-080842.1 & 3.171 & 9.13 & 12.29 & 46.97 & 13.89 & 13.26 & \nodata \\
99 & SDSS 234625.66-001600.4 & 3.490 & 9.51 & 13.34 & 47.08 & 14.00 & 14.13 & 10.24 \\
100 & SDSS 235718.36+004350.4 & 4.364 & 7.42 & 11.32 & 47.16 & 14.08 & 13.40 & 8.69 \\

\enddata

\tablenotetext{a}{Optical redshift from version 7 of the SDSS quasar catalog \citep{sch10}.}
\tablenotetext{b}{Observed flux density at 22 \um from the WISE All Sky Catalog available at http://wise2.ipac.caltech.edu/docs/release/allsky/.  Zero point of 22 \um magnitude listed in catalog taken as 8284 mJy; typical uncertainties for sources with fluxes listed are $\pm$ 15 $\%$.  }
\tablenotetext{c}{Flux density $f_{\nu}$(7.8 \ums) at observed wavelength corresponding to rest wavelength 7.8 \ums, determined by scaling $f_{\nu}$ (observed 22 \ums) to $f_{\nu}$(rest frame 7.8 \ums) using tabulated redshift and template spectrum shown in Figure 6.} 
\tablenotetext{d}{Rest frame luminosity $\nu L_{\nu}$(7.8 \ums) in erg s$^{-1}$ determined as $\nu L_{\nu}$(7.8 \ums) =  4$\pi$D$_{L}$$^{2}$[$\nu$/(1+z)]$f_{\nu}$(7.8 \ums), for $\nu$ corresponding to 7.8 \ums, taking luminosity distances from \citet{wri06}:  http://www.astro.ucla.edu/~wright/CosmoCalc.html, for H$_0$ = 74 \kmsMpc, $\Omega_{M}$=0.27 and $\Omega_{\Lambda}$=0.73. }
\tablenotetext{e}{Total infrared luminosity $L_{IR}$ determined using calibration log [$L_{IR}$/$\nu L_{\nu}$(7.8 \ums)] = 0.51 from \citet{sar11} (Log [$\nu L_{\nu}$(\ldot)] = log [$\nu L_{\nu}$(erg s$^{-1}$)] - 33.59.)} 
\tablenotetext{f} {Bolometric luminosity $L_{bol}$ from \citet{she11}, adopting a scaling of $L_{bol}$ = 5.15$\lambda L_{\lambda}$(0.30 \ums) for z $<$ 1.9 and $L_{bol}$ = 3.81$\lambda L_{\lambda}$(0.135 \ums) for z $>$ 1.9. }
\tablenotetext{g}{Virial black hole mass from width of CIV line estimated by \citet{she11}. }
\tablenotetext{h}{Gravitationally lensed source from \citet{ina10} and references therein.}

\end{deluxetable}

\subsection{Comparisons of Infrared Luminosity to Black Hole Masses and Bolometric Luminosities}

The comparisons of mid-infrared luminosity $\nu L_{\nu}$(7.8 \ums) with virial black hole mass are shown in Figure 7.   For the local AGN with virial black hole masses from reverberation mapping (crosses and diagonal fit), the luminosity increases linearly with black hole mass (BHM). The linear fit which is shown is log $\nu L_{\nu}$(7.8 \ums) = 37.2 + (0.874 $\pm$ 0.11)log BHM for luminosity in erg s$^{-1}$ and black hole mass in M$_{\odot}$.  The dispersion about the fit in log $\nu L_{\nu}$(7.8 \ums) of $\pm$ 0.5 indicates that this relation is reliable for local type 1 AGN to a factor of 3.  

The dispersion can be caused by many factors: uncertainty in BHM estimates, differences in accretion efficiency, and differences in ratios of mid-infrared luminosity to primary accretion-derived luminosity.  The latter factor is already known to have a cosmic dispersion of $\pm$ 0.35 in dex from Figure 4, which compares infrared dust luminosity with intrinsic hard X-ray luminosity.   Taking out this dispersion from the $\pm$ 0.5 dispersion in Figure 7 leaves a cosmic dispersion of 0.36 in dex (scatter of a factor of 2.3) for the remaining factors of black hole mass uncertainty and accretion efficiency for the local type 1 AGN used for reverberation mapping.  

\begin{figure}

\figurenum{7}
\includegraphics[scale=0.8]{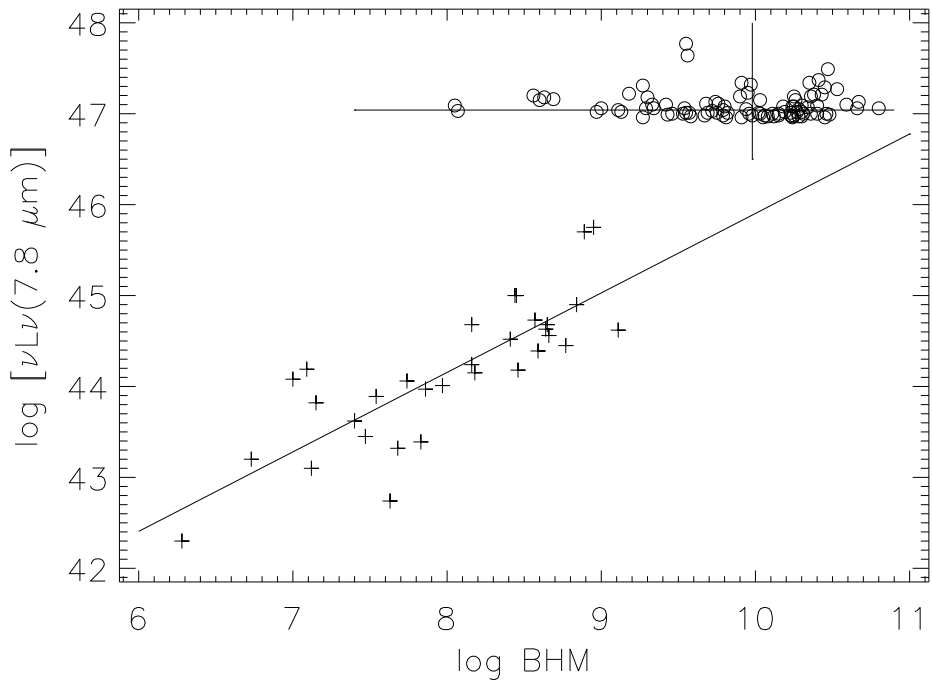}
\caption{Comparison of $\nu L_{\nu}$(7.8 \ums) in erg s$^{-1}$ with virial black hole mass (solar masses), determined using emission lines from broad line region.  Crosses are sources from local AGN with reverberation mapping, determined with Balmer line widths \citep{pet04}. Circles are SDSS quasars, for which BHM estimated from CIV emission line widths \citep{she11}. The diagonal line is the fit to the local AGN.  Large cross shows the medians for the SDSS quasars. }

\end{figure}

The fit shown in Figure 7 for local type 1 AGN does not, however, extrapolate to the luminous SDSS/WISE quasars even though all of these are spectroscopically type 1 quasars (based on broad permitted emission lines).  In the most extreme case, one quasar is overluminous by a factor of 1000 compared to the local AGN relation between BHM and mid-infrared luminosity!  The median values in BHM and luminosity for the quasars show infrared luminosities about a factor of 10 larger than would be expected from the virial black hole mass (or, conversely, a BHM underestimated by a factor of 10 compared to the luminosity).  Although this difference is large, it is only about a 2$\sigma$ effect compared to the uncertainties in the virial mass determinations \citep{pet04,she11}. 

There are various reasons that can explain this lack of fit for luminous quasars compared to local AGN.  The quasars have been selected as those SDSS quasars that are the most luminous in the mid-infrared as determined from WISE, unlike any selection applied for the local AGN.  Selecting for large values of $\nu L_{\nu}$(7.8 \ums) favors the type 1 sources with the most extreme hot dust emission.  This extreme could arise for a combination of reasons involving the geometry of the dust, including hotter dust because the dust is closer to the AGN or more dust emission because the dust covering factor is greater.  These possibilities are considered further in the next section. 

\subsection{Infrared Luminosities compared to Bolometric Luminosities}

The primary bolometric luminosity $L_{bol}$ arising from the AGN is the fundamental luminosity measurement that is ultimately needed.  In the absence of extinction, this $L_{bol}$ can be scaled from rest frame optical or ultraviolet observations using multiwavelength templates, and such measurements are the primary sources of $L_{bol}$ \citep{she11}.  To understand the uncertainties of such estimates for $L_{bol}$, we compare to the independent measures of infrared luminosity. 

The objective in using an infrared luminosity measure is to achieve an independent measurement of total luminosity that avoids template assumptions and is not sensitive to extinction effects, so that all AGN and quasars can be uniformly compared without concerns about classification or orientation.  The simple explanation assumed for the total infrared luminosity $L_{IR}$ is that primary radiation at optical through soft X-ray wavelengths has been absorbed by dust and then reradiated from the heated dust as $L_{IR}$ via the dust continuum emission.  If all of the emitting dust is optically thick to the shorter wavelength primary radiation, then the total infrared luminosity emitted must equal the total primary luminosity absorbed.  If the dust covering factor is large, therefore, $L_{IR}$ $\sim$ $L_{bol}$. 

The total infrared luminosity $L_{IR}$ is the integrated luminosity from 5 \um to 1000 \um as defined by \citet{san96} using infrared fluxes in the photometric bands of the Infrared Astronomical Satellite (IRAS). The $L_{IR}$ have been empirically scaled to $\nu L_{\nu}$(7.8 \ums) by \citet{sar11} using type 1 AGN having measures of both $L_{IR}$ from IRAS and $\nu L_{\nu}$(7.8 \ums) from IRS spectra; the scaling is log [$L_{IR}$/$\nu L_{\nu}$(7.8 \ums)] = 0.51 $\pm$ 0.21.  We adopt this scaling to estimate $L_{IR}$ for the local AGN in Table 2 and the SDSS/WISE quasars in Table 3.  

For the local AGN in Table 2, the $L_{bol}$ are determined from \citet{ric06} as $L_{bol}$ = 9.26$\lambda L_{\lambda}$(0.51 \ums) using the $\lambda L_{\lambda}$(0.51 \ums) for the AGN component only (excluding the galactic starlight component) from \citet{ben09}.   For the SDSS/WISE quasars in Table 3, the $L_{bol}$ are taken from \citet{she11}, adopting a scaling of $L_{bol}$ = 5.15$\lambda L_{\lambda}$(0.30 \ums) for z $<$ 1.9 and $L_{bol}$ = 3.81$\lambda L_{\lambda}$(0.135 \ums) for z $>$ 1.9.   The references cited review the large uncertainties in $L_{bol}$, especially because no extinction corrections are made to the optical or ultraviolet fluxes to obtain $L_{bol}$.  Our goal is only to illustrate an overall comparison of systematic differences between the independent observational measures $L_{IR}$ and $L_{bol}$, so we do not examine this comparison for individual sources.   

The comparisons of $L_{IR}$ and $L_{bol}$ are shown in Figure 8.   For the AGN sample (crosses in Figure 8), the median log $L_{IR}$/$L_{bol}$ = -0.35.  This is the systematic difference that arises when using these two independent measures of total luminosity.  If the optically determined $L_{bol}$ are not affected by extinction, i.e. the observed optical luminosity at rest frame 0.51 \um for type 1 AGN shines past dust clouds with no interruption, then the ratio $L_{IR}$/$L_{bol}$ is also a measure of the fraction of primary radiation intercepted by dust clouds in directions outside the observer's line of sight.  This ratio is equivalent to the covering factor for the dust clouds, corresponding to 45\% with this value of the median. This value of 45\% is similar to the 55\% dust covering factor for BAT AGN determined by silicate absorption in the IRS spectra, discussed in section 2.3.  

\begin{figure}
\figurenum{8}
\includegraphics[scale=0.8]{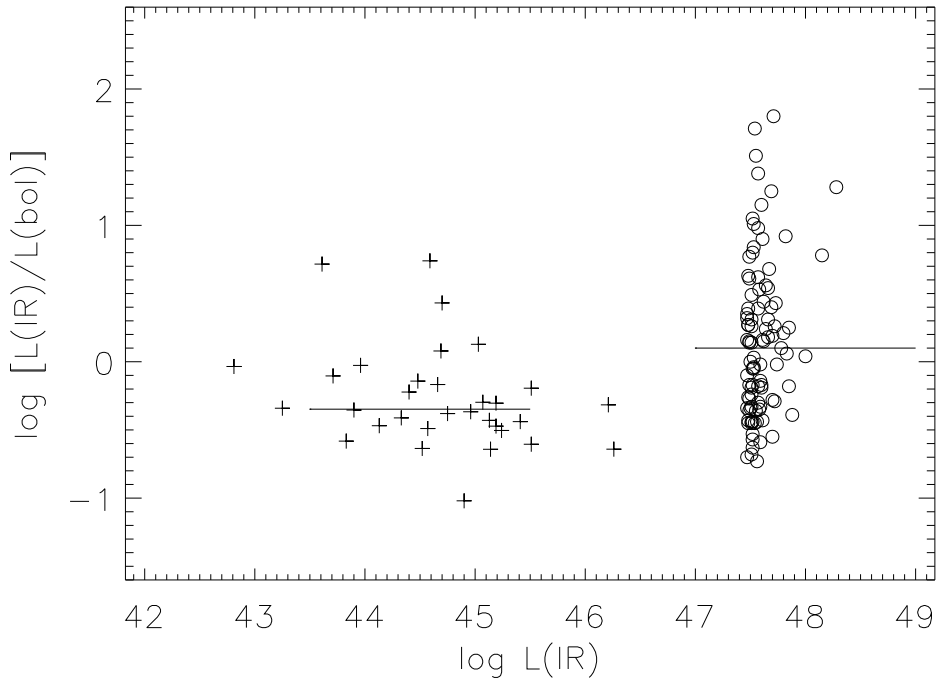}
\caption{Comparison of total infrared luminosities $L_{IR}$ in erg s$^{-1}$ to total bolometric luminosity $L_{bol}$ scaled from observed optical or rest frame ultraviolet luminosity, determined as described in the text.  Crosses are local AGN in Table 2 with black hole masses from reverberation mapping.  Circles are the 97 SDSS/WISE quasars with z $>$ 1.5 which are the most luminous in $\nu L_{\nu}$(7.8 \ums), omitting gravitationally lensed sources (Table 3).  Horizontal lines are medians for the two samples.} 

\end{figure}

The ratio $L_{IR}$/$L_{bol}$ for the most infrared-luminous SDSS/WISE quasars (Table 3 and circles in Figure 8) has a median of log $L_{IR}$/$L_{bol}$ = 0.1, indicating that the optically-derived and infrared-derived measures of total luminosity are nearly the same systematically.  This implies a nearly 100\% dust covering factor, which can be explained for these type 1 quasars only if the measured $L_{bol}$ derives from ultraviolet luminosity that just happens to be observed without extinction through small holes between dust clouds. 

The ratio extends to values of log $L_{IR}$/$L_{bol}$ $\sim$ 2, however, indicating that for some type 1 quasars, the infrared measures can give much larger total luminosities than the $L_{bol}$ measures derived from rest frame ultraviolet luminosity.  If log $L_{IR}$/$L_{bol}$ $>$ 0, the simple equating of $L_{IR}$/$L_{bol}$ to a dust covering factor cannot apply; this ratio cannot exceed unity when all primary radiation is absorbed and the covering factor is 100\%.  Cases with $L_{IR}$ $>$ $L_{bol}$  would anyway be nonsensical within our assumptions, because no rest frame optical or ultraviolet radiation could be observed from which to derive an $L_{bol}$ if optically thick dust covers 100\% of the source.  

Extreme values of $L_{IR}$ $>$ $L_{bol}$ can be qualitatively explained in part by selection effects that arise from choosing only those type 1 quasars which are most luminous in the mid-infrared.  There is a one sigma dispersion of $\pm$ 0.2 in the ratio log $L_{IR}$/$\nu L_{\nu}$(7.8 \ums) calibrated empirically by \citet{sar11}.  Furthermore, section 2.3 showed a dispersion in log of $\pm$ 0.35 for the ratio of hard X-ray flux to mid-infrared $\nu f_{\nu}$(7.8 \ums).  These two factors reflect the cosmic dispersion in dust temperature and/or intrinsic spectral energy distributions that affect the fundamental relation between $L_{IR}$ and the intrinsic bolometric luminosity of the AGN.  The convolved one sigma dispersion of $\pm$ 0.4 can account for much of the apparent excess of $L_{IR}$ for those quasars which are most luminous in $\nu L_{\nu}$(7.8 \ums); the most extreme source in $L_{IR}$/$L_{bol}$ is only 4 sigma from the median.  Variability in the ultraviolet luminosity that is used to measure $L_{bol}$ could also explain some unusual ratios because the $L_{IR}$ arising from dust reradiation should not be variable.

Furthermore, those quasars with the largest dust covering factors would also be those most likely to have some dust extinction affecting the rest frame 
ultraviolet that leads to $L_{bol}$.  Measuring $L_{bol}$ too small because the ultraviolet is not corrected for extinction would also lead to erroneously large values of $L_{IR}$/$L_{bol}$, and this effect is the most straightforward explanation for sources with large values of $L_{IR}$/$L_{bol}$.  

The most important conclusion from Figure 8 is that infrared measures can give substantially larger values of total luminosity than do $L_{bol}$ measures derived from rest-frame ultraviolet luminosities, even for type 1 quasars.  This means that infrared measures are essential to derive total luminosities not only for the very dusty, optically obscured DOGs, but also for the most luminous optically discovered quasars.  

\section{Maximum Luminosities for Dusty Quasars}

A compelling observational challenge is finding the most luminous sources in the universe and understanding how they have changed with cosmic time.  What are the maximum luminosities?  Has the epoch of maximum luminosity been observed?  How do different classes of sources compare in luminosity?  Our ultimate goal is to compare luminosities for all sources, including both the classical type 1 quasars and the extremely dusty, optically obscured DOGs.  Making this comparison using observed infrared luminosities $\nu L_{\nu}$(7.8 \ums) allows a comparison of the most luminous sources among all categories regardless of extinction or discovery technique and independent of assumed multiwavelength spectral templates ("K corrections").

\subsection{Mid Infrared Dust Luminosities of Type 1 AGN and Quasars}

The BAT and SDSS/WISE sources are ideal samples for comparing luminosities of type 1, silicate emission sources between the local and high redshift universe because both samples arise from large survey areas (BAT covers the full sky, and the SDSS quasar catalog covers 9400 deg$^{2}$). The luminosity scaling with redshift of the silicate emission BAT AGN (all sources in Table 1 with silicate strength $>$ 1) and the 97 SDSS/WISE quasars from Table 3 (not including the 3 lensed quasars) is shown in Figure 9. 

This Figure displays only the 97 quasars most luminous in $\nu L_{\nu}$(7.8 \ums) from the sample of 16242 SDSS quasars in the interval 1.5 $<$ z $<$ 5 with WISE 22 \um detections.  If the sample included all WISE detections ($f_{\nu}$(22 \ums) $>$ 3 mJy), the luminosity detection limit would be as shown by the thick solid line in Figure 9.  All of the remaining 16145 SDSS/WISE quasars with 22 \um detections would fill in above this limit line. 

\begin{figure}
\figurenum{9}
\includegraphics[scale=0.80]{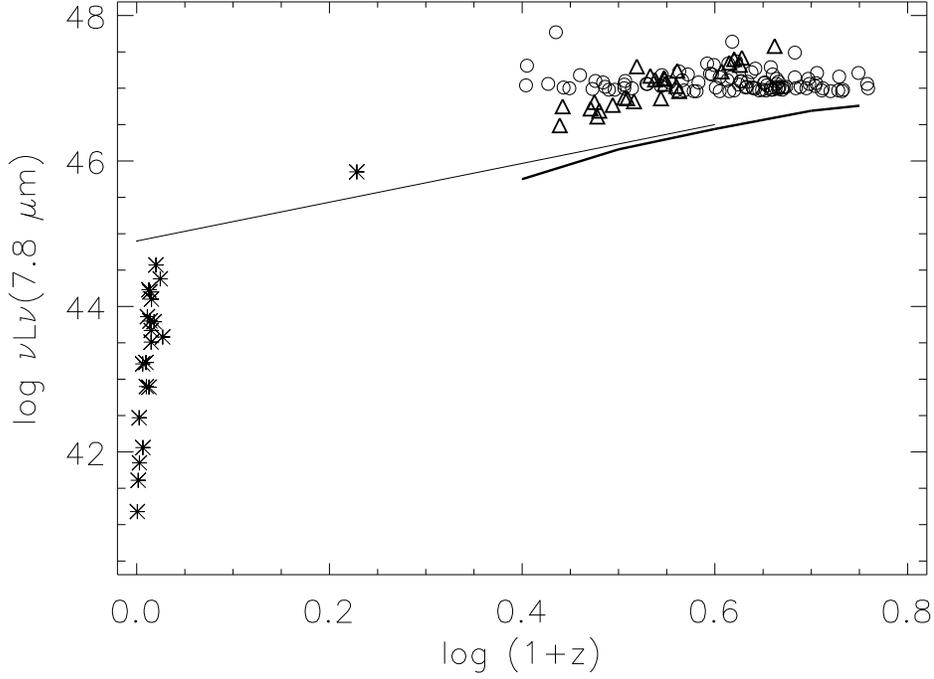}
\caption{Continuum luminosity $\nu L_{\nu}$(7.8 \ums) in erg s$^{-1}$ compared to redshift for dusty AGN and quasars. Asterisks are silicate emission BAT sources in Table 1 (isolated BAT source in center is 3C 273); circles are SDSS/WISE quasars in Table 3 (excluding 3 lensed sources). Triangles are WISE sources selected for heavy dust obscuration \citep{wu12} scaled to 7.8 \um luminosity as described in text. Thin solid line is envelope of most luminous silicate absorbed ULIRGs or DOGs with 0 $<$ z $<$ 3 measured from $Spitzer$ IRS spectra in \citet{wee09}. Thick solid line is faint luminosity limit (lower envelope) found for all SDSS/WISE sources in interval 1.5 $<$ z $<$ 5 using all 16242 SDSS/WISE 22 \um detections brighter than 3 mJy combined with template in Figure 6. (Log [$\nu L_{\nu}$(7.8 \ums)(\ldot)] = log [$\nu L_{\nu}$(7.8 \ums)(erg s$^{-1}$)] - 33.59.)} 


\end{figure}

The results shown in Figure 9 are empirical, observed scaling of luminosities with redshift.  These results allow determination of the maximum infrared luminosities which have been discovered so far and the epoch at which they are found.  The scaling of luminosities with redshift does not describe "luminosity evolution" as normally defined, because only the maximum observed luminosities are traced as a function of redshift.  Luminosity functions and space densities must be considered in order to determine actual luminosity evolution.   

Despite the limitations of a simplified interpretation, several interesting results are obvious in Figure 9.  The range of observed luminosities measured with the single observed parameter $\nu L_{\nu}$(7.8 \ums) covers a factor exceeding 10$^{6}$.  (We remind again that the lower luminosity cutoff for SDSS/WISE quasars is artificial; we selected only the 100 most luminous sources.)  The most luminous SDSS/WISE quasars have essentially common maximum luminosities for all z $>$ 1.5, with maximum luminosities reaching log $\nu L_{\nu}$(7.8 \ums) = 47.5, corresponding to log $L_{IR}$ = 48 (erg s$^{-1}$) with our empirical $L_{IR}$/$\nu L_{\nu}$(7.8 \ums) calibration, or $L_{IR}$ = 10$^{14.4}$ \ldot!  

There is no indication of any particular redshift at which the luminosity achieves a maximum, and the maximum maintains to the highest observed redshift.  A crucial future question for the observer is determining to which redshifts these extremely luminous sources continue to exist.  

\subsection{Mid Infrared Dust Luminosities of Optically Obscured Quasars}

One of our primary objectives is to determine if the most luminous quasars have been overlooked in optical surveys because they are DOGs with heavy extinction in the rest frame ultraviolet.  A comparison of the most luminous known DOGs with the type 1 quasars is also given in Figure 9.  

The envelope of maximum luminosities $\nu L_{\nu}$(7.8 \ums) for silicate absorbed local ULIRGs and high redshift DOGs having silicate strength $<$ 0.5 which have been measured with IRS spectra to z = 3 is shown as the line in Figure 9 \citep{wee09}. Also shown for comparison is a new set of optically faint quasars discovered in the all sky WISE survey using infrared photometric color criteria to select obscured sources \citep{eis12,wu12}. These sources have optical redshifts 1.7 $<$ z $<$ 3.6 and are interpreted as the most luminous DOGs discovered in the WISE all sky sample.

The comparison shows that the most luminous type 1 SDSS/WISE quasars exceed the dust luminosity of the most luminous, heavily obscured $Spitzer$ DOGs by a factor of $\sim$ 5 (at z = 3, the redshift limit of currently measured DOGs).  Much smaller areas have been surveyed by $Spitzer$ to find DOGs, however, so space density comparisons accounting for survey volumes may infer that equally or more luminous DOGS could be found in wide area surveys.  Such hyperluminous DOGs should be detected within the WISE survey; the WISE luminosity limit line in Figure 9 for 1.5 $<$ z $<$ 5 illustrates that DOGS which are comparable to the most luminous DOGs already discovered with $Spitzer$ would be readily detectable by WISE.  The impossibility of new mid-infrared spectra means, however, that candidate WISE DOGs cannot be unambiguously compared to the $Spitzer$ DOGs whose redshifts arise from infrared spectra and whose identification is completely independent of optical measurements.    

Because the infrared spectra of these WISE DOGs are not known, there is no confident template with which to relate observed frame 22 \um to rest frame 7.8 \ums.  Mid-infrared spectra could range from the type 1 template in Figure 6 to the absorbed AGN spectrum in Figure 1.  A larger 7.8 \um luminosity arises if the silicate absorbed spectrum in Figure 1 is the template because observed-frame 22 \um observations then require larger corrections to rest-frame 7.8 \um (observed-frame 22 \um is $<$ 7.8 \um rest frame for all z $>$ 1.8).  We estimate in this way the maximum luminosities for the 26 WISE DOGs in \citet{wu12} with redshifts and 22 \um photometry. These are the luminosities shown in Figure 9 (the most luminous with these assumptions is W0410-0913).   

Overall, the most luminous of these WISE DOGs are similar to the most luminous SDSS/WISE quasars in $\nu L_{\nu}$(7.8 \ums), but the DOGs must be more obscured because they are too faint optically to be within the SDSS.  Many WISE DOGs are optically brighter than the $Spitzer$ DOGs, however, indicating that they are not as obscured when judged by the infrared/optical flux ratios.  In the absence of infrared spectra, we cannot conclude if the WISE DOGs have silicate absorption as strong as the $Spitzer$ DOGs, or represent an intermediate category with less silicate absorption. An indication that the WISE DOGs and $Spitzer$ DOGs represent similar populations of obscured AGN is the similar slope for increasing luminosity with increasing redshift, differing only that the WISE sources are systematically more luminous by about a factor of 3 (0.5 in dex) compared to the most luminous $Spitzer$ sources.  

Figure 9 summarizes what is known observationally regarding the most infrared luminous quasars.  At present, the optically discovered SDSS/WISE quasars remain the most luminous dusty sources known in the universe.  We find no evidence of hyperluminous, obscured quasars whose infrared luminosities significantly exceed the infrared luminosities of optically discovered quasars.  This result addresses only the question of maximum quasar luminosities.  It is also important to compare the total space densities of optically discoverable, less obscured quasars ($\sim$ type 1) to the optically faint, heavily obscured quasars ($\sim$ type 2), and to understand evolution for these different populations.  The similar trends among both WISE DOGs and $Spitzer$ DOGs for maximum luminosity to increase with redshift, contrasted to constant maximum luminosity for type 1 quasars, is an observational hint that dusty populations may become more important at high redshift. 

There are no wide area samples of heavily obscured DOGs with spectroscopic infrared classifications, but there are limited area samples of heavily absorbed DOGs with $Spitzer$ IRS spectra.  An initial summary of these silicate absorbed quasars was in \citet{wee09}; we have now identified 74 sources in CASSIS with 1.5 $<$ z $<$ 3.35 and silicate strength $\la$ 0.5.  In a future analysis, we will compare space densities at similar redshifts among the heavily absorbed DOGs already discovered by $Spitzer$, the all sky WISE DOGs, and the SDSS/WISE type 1 quasars.  This will allow more quantitative conclusions about the dusty fraction of the most luminous quasars and whether dust content evolves with redshift.  Such an analysis will also allow predictions of how dusty quasars at even higher redshifts should appear in far infrared and submillimeter surveys, depending on various assumptions of how the luminosity function at z $\sim$ 2.5 extrapolates to higher redshifts.


\section{Summary and Conclusions}

Mid-infrared spectroscopic results from $Spitzer$ IRS spectra are given for 125 hard X-ray AGN (14-195 keV) from the $Swift$ BAT sample, calibrating strength of the 9.7 \um silicate absorption or emission feature compared to X-ray luminosity and optical classification.  The presence of silicate emission or absorption defines an infrared AGN classification describing whether or not AGN are observed through dust clouds, and the quantitative silicate strength is found to be consistent overall with optical classifications from type 1 through type 2.  The silicate classification indicates that 55\% of the BAT AGN are observed through dust clouds. 

From the BAT sample, the mid-infrared dust continuum luminosity is found to be an excellent indicator of intrinsic AGN luminosity, scaling closely with the hard X-ray luminosity, log [$\nu L_{\nu}$(7.8 \ums)/$L$(X)] = -0.31 $\pm$ 0.35.  This scaling is independent of classification determined from silicate strength, demonstrating that $\nu L_{\nu}$(7.8 \ums) is a consistent indicator of luminosity that applies equally to unobscured, type 1 AGN as well as highly obscured type 2 AGN.

IRS spectra are also presented for 32 AGN with black hole mass measures from reverberation mapping.  Mid-infrared dust luminosity scales closely with black hole mass, as log $\nu L_{\nu}$(7.8 \ums) = 37.2($\pm$ 0.5) + 0.87 log BHM for luminosity in erg s$^{-1}$ and BHM in M$_{\odot}$.  

The 100 most luminous type 1 quasars as measured in $\nu L_{\nu}$(7.8 \ums) are found by comparing SDSS optically discovered quasars with WISE photometry at 22 \ums, using an empirical template determined from IRS spectra of type 1 sources for scaling to rest frame 7.8 \ums.  The most luminous SDSS/WISE quasars have similar maximum infrared luminosities for all 1.5 $<$ z $<$ 5, reaching log $\nu L_{\nu}$(7.8 \ums) = 47.5 (erg s$^{-1}$).  With our empirical scaling for type 1 sources that log $L_{IR}$/$\nu L_{\nu}$(7.8 \ums) = 0.51, for $L_{IR}$ the total infrared luminosity from 5 \um to 1000 \ums, this maximum is $L_{IR}$ = 10$^{14.4}$ \ldot.  There is no indication of any particular redshift for z $<$ 5 at which the infrared luminosity is maximum.

For the black hole AGN sample and SDSS/WISE quasars, the bolometric luminosities $L_{bol}$ estimated by scaling from rest frame optical or ultraviolet luminosities are compared to total infrared luminosities $L_{IR}$.  For the local AGN, the median log $L_{IR}$/$L_{bol}$ = -0.35, indicating a covering factor of 45\% for the dust clouds that absorb intrinsic $L_{bol}$ and reemit as $L_{IR}$.  For the SDSS/WISE quasars deliberately selected as the most luminous infrared sources, the median log $L_{IR}$/$L_{bol}$ = 0.1, with extremes reaching values of $\sim$ 2, indicating that ultraviolet-derived $L_{bol}$ can be seriously underestimated even for type 1 quasars. 

The optically discovered SDSS/WISE type 1 quasars are the most luminous dusty sources currently known in the universe.  We find no evidence of hyperluminous, obscured quasars whose infrared luminosities significantly exceed the infrared luminosities of optically discovered quasars. The most luminous type 1 quasars exceed the dust luminosity of the most luminous, heavily obscured quasars (the DOGs found in limited area surveys with $Spitzer$) by a factor of $\sim$ 5 and have comparable maximum luminosities to the recently discovered population of optically faint, dusty quasars from the all sky WISE survey.

\acknowledgments

This work is based in part on observations made with the
$Spitzer$ Space Telescope, which is operated by the Jet Propulsion
Laboratory, California Institute of Technology, funded by the National Aeronautics and Space Administration. 
Support for this work by the IRS GTO team at Cornell University was provided by NASA through Contracts
issued by JPL/Caltech. 

This publication makes use of data products from the Wide-field Infrared Survey Explorer, which is a joint project of the University of California, Los Angeles, and JPL/Caltech, funded by NASA.

We also acknowledge data products from the SDSS.  Funding for the SDSS has been provided by the Alfred P. Sloan Foundation, the Participating Institutions, the National Science Foundation, the U.S. Department of Energy, NASA, the Japanese Monbukagakusho, the Max Planck Society, and the Higher Education Funding Council for England. The SDSS is managed by the Astrophysical Research Consortium for the Participating Institutions: the American Museum of Natural History, Astrophysical Institute Potsdam, University of Basel, University of Cambridge, Case Western Reserve University, University of Chicago, Drexel University, Fermilab, the Institute for Advanced Study, the Japan Participation Group, Johns Hopkins University, the Joint Institute for Nuclear Astrophysics, the Kavli Institute for Particle Astrophysics and Cosmology, the Korean Scientist Group, the Chinese Academy of Sciences (LAMOST), Los Alamos National Laboratory, the Max-Planck-Institute for Astronomy (MPIA), the Max-Planck-Institute for Astrophysics (MPA), New Mexico State University, Ohio State University, University of Pittsburgh, University of Portsmouth, Princeton University, the United States Naval Observatory, and the University of Washington.

\end{document}